\documentclass{aa}
\usepackage{graphicx}
\usepackage{txfonts}
\usepackage{multirow} 
\usepackage[outdir=./]{epstopdf}
\usepackage{diagbox} 
\usepackage{setspace}
\usepackage{float} 
\usepackage[usenames,dvipsnames,svgnames,table]{xcolor} 
\usepackage[colorlinks,allcolors=blue]{hyperref}
\begin{document}

   \title{Very fast variations of SiO maser emission in evolved stars}
        
   \author{M. Gómez-Garrido\inst{1,2}, V. Bujarrabal\inst{1}, J. Alcolea\inst{3}, R. Soria-Ruiz\inst{3}, P. de Vicente\inst{2} \and J.-F. Desmurs\inst{3}}

   \institute{Observatorio Astronómico Nacional (OAN-IGN), Apartado 112, E-28803 Alcalá de Henares, Spain\\email:{ m.gomezgarrido@oan.es}   
         \and
         Centro de Desarrollos Tecnológicos, Observatorio de Yebes (IGN), 19141 Yebes, Guadalajara, Spain    
        \and
        Observatorio Astron\'omico Nacional (OAN-IGN), Alfonso XII 3, E-28014, Madrid, Spain
 }

\authorrunning{Gómez-Garrido, M. et al. }

   \date{Accepted 08 September 2020} 

 
  \abstract
   {Stars on the asymptotic giant branch (AGB) are long-period variables that present strong flux variations at almost all wavelengths, including the SiO maser lines. The periods of these variations are of 300–500 days in Mira-type stars and somewhat shorter in semi-regular variables. The variability of the SiO lines on short timescales has been investigated, but the data are inconclusive.}
   {We aim to study the time evolution of the SiO maser lines in Mira-type and semi-regular variables at short timescales. We also discuss the origin of the observed fast variations.}
   {We observed the SiO maser lines at 7\,mm ($^{28}$SiO $v$=1,2 $J$=1--0) and 3\,mm ($^{28}$SiO $v$=1 $J$=2--1) using the 40\,m Yebes antenna and the 30\,m IRAM telescope, respectively, with a minimum spacing of 1 day. We studied the semi-regular variables RX\,Boo and RT\,Vir and the Mira-type variables U\,Her, R\,LMi, R\,Leo, and $\chi$\,Cyg. We performed a detailed statistical analysis of the variations on different timescales.}
   {RX\,Boo shows strong and fast variations in the intensity of the different spectral features of the SiO lines at 7\,mm and 3\,mm. On a timescale of one day, we find variations of $\gtrsim$ 10\% in 25\% of the cases. Variations of greater than $\sim$50\% are often found when the observations are separated by 2 or 3 days. A similar variation rate of the SiO lines at 7\,mm is found for RT\,Vir, but the observations of this object are less complete. On the contrary, the variations of the SiO maser line intensity in the Mira-type variables are moderate, with typical variation rates around $\lesssim$\,10\% in 7 days. This phenomenon can be explained by the presence of particularly small maser-emitting clumps in semi-regular variables, which would lead to a strong dependence of the intensity on the density variations due to the passage of shocks.} 
   {}


   \maketitle
%

\section{Introduction}\label{intro}

During the asymptotic giant branch (AGB) phase, strong pulsations and mass loss take place, and a circumstellar envelope (CE) is formed around the star as a result. The properties of the inner part of the CE, a particularly important region to understand the mass-loss process, can be studied using SiO maser lines {\citep[e.g.,][]{habing96}. SiO maser emission takes place between rotational levels ($J$=1–0, 2–1, 3–2, ...) in excited vibrational states ($v$=1, 2, 3, and 4). The most intense maser lines are $v$=1,2 $J$=1-0 and $v$=1 $J$=2–1, with the $v$=2 $J$=2–1 line usually not detected \citep{pardo98}. The rotational lines in the vibrational state $v$=0 are often much weaker than the $v$\,$>$\,0 maser lines and can also present maser emission, particularly in the $^{29}$SiO and $^{30}$SiO isotopes. $^{28}$SiO $v$=0 lines often show relatively weak maser features \citep{bobcla04,devicente16}, with the rest of the profile being due to thermal emission. Very long baseline interferometry (VLBI) studies show that SiO masers come from many compact spots (a few milliarcseconds wide), which form ring-like structures at typical distances of a few stellar radii \citep[see][]{diamond94,soriaruiz04}.

Stars on the AGB are long-period variables that present strong flux variations at optical, infrared, and radio wavelengths, including the SiO maser lines. The periods of these variations are of 300-500 days in Mira-type stars, which are relatively regular variables. Typically, SiO masers show significant variations ($\sim$20\%) over 1-3 months. On the contrary, the variability of the $v$=0 lines is negligible; at least the thermal component of the profiles is constant. Semi-regular and irregular variables are characterized by a less well-defined period and smaller amplitudes \citep{habing96}. In regular variables, the SiO maser variability follows the infrared curve, with a phase lag of 0.1-0.2 with respect to the optical curve. This behavior strongly suggests that the maser pumping is mainly radiative through the absorption of stellar 8\,$\mu m$ photons \citep[e.g.,][]{pardo04,soriaruiz04}. However, other pumping mechanisms have been proposed to explain the observed properties of SiO maser in AGB stars, such as for example in the collisional models \citep[e.g.,][]{locket92,humphreys02}. Semi-regular variables present a strong and erratic variability in the maser line intensity on timescales of several months. At the minimum of their variability curves, SiO lines are weak and sometimes not detected for months in semi-regular variables. Previous studies of variable stars only included a few semi-regular variables, such as for example RT\,Vir in \cite{alcolea99} and \cite{pardo04}, where the 7\,mm $^{28}$SiO masers were observed. \cite{kim08} also performed VLBI observations at 7\,mm of the semi-regular R\,Crt in three epochs (May 21, 2015, Jan 07, 2016, and Jan 26, 2016). In general, the SiO emission from semi-regular variables was found to be weak and its variability erratic. Some indications of the presence of fast variations were reported in these previous studies, but they were not tight and systematic enough to characterize fast variability in semi-regular variables. 

\begin{table*}
\caption{Observed sources and epochs.}

\small
\begin{tabular}{ l c c c c c c c c}
\hline 
\noalign{\smallskip}
Source & A.R. & dec. &  Var. &  \multicolumn{5}{c}{Observation dates}\\ 
 & (h:m:s) & (d:m:s) & type & Aug 2016 & Sep 2016 & Mar 2017 & Aug-Sep 2017 & Aug-Sep 2017 (30\,m)\\
\hline
 \noalign{\smallskip}
\multirow{2}{*}{RX Boo} & \multirow{2}{*}{14:24:11.63} & \multirow{2}{*}{$+$25:42:13.4}  & \multirow{2}{*}{Semireg.}\ & 10,11,12,13,14,15 & 10,14,15,17 & 12,16,17 & 23,24,25,26,27,28 & 23,24,25,26,27 \\  
 & & & & 16,17,18,20,21 & 18,19,20 & 18,20,22 & 29,30,31,01,02 &  29,31,01\\ \noalign{\smallskip}
\multirow{2}{*}{U Her} & \multirow{2}{*}{16:25:47.47} & \multirow{2}{*}{$+$18:53:32.9} & \multirow{2}{*}{Mira} & \multirow{2}{*}{not obs.} & \multirow{2}{*}{not obs.} & 16,17,18  & 23,24,25,27,28  & 23,24,26,31,01\\
&  &  &  &  &  & 20,22 & 29,30,31,01,02 & \\ \noalign{\smallskip} 
\multirow{2}{*}{R Leo} & \multirow{2}{*}{09:47:33.49} & \multirow{2}{*}{$+$11:25:43.7} & \multirow{2}{*}{Mira} & \multirow{2}{*}{not obs.} & \multirow{2}{*}{not obs.} & 11,15,16 & \multirow{2}{*}{not obs.}  & \multirow{2}{*}{not obs.}  \\ 
 & & &  & & & 17,19,21 &  &\\ \noalign{\smallskip}
\multirow{2}{*}{R LMi} & \multirow{2}{*}{09:45:34.28} &  \multirow{2}{*}{$+$34:30:42.8} & \multirow{2}{*}{Mira} & \multirow{2}{*}{not obs.} & \multirow{2}{*}{not obs.} & 11,15,16 & \multirow{2}{*}{not obs.}  & \multirow{2}{*}{not obs.} \\ 
 & &  & & & & 17,19,21 &  &\\ \noalign{\smallskip} 
\multirow{2}{*}{$\chi$ Cyg} & \multirow{2}{*}{19:50:33.92} & \multirow{2}{*}{$+$32:54:50.6} & \multirow{2}{*}{S-type} & \multirow{2}{*}{not obs.} & \multirow{2}{*}{not obs.}  & 16,17,18 & 23,25,27,28 & 23,24,26,31,01 \\ 
 &  & &  &  &  & 20,22 & 30,01,02 &  \\ \noalign{\smallskip} 
 \hline\noalign{\smallskip} 
RT Vir & 13:02:37.98 & $+$05:11:08.4 & Semireg.\ & \multicolumn{5}{c}{Observation dates: 7, 8, 9, 11, 13, 16 Nov. 2017} \\ \noalign{\smallskip}
\hline  

\end{tabular} 
\label{tab.sourcesandobservations}
\end{table*}

\cite{pijpers94} investigated the variability of the $^{28}$SiO $v$=1 $J$=1--0 line on short timescales in the Mira-type variables R\,Leo and R\,Cas, in the SRc variable VY\,CMa, and in the star-forming region Ori\,A. These latter authors found significant changes in the total emission of the SiO maser lines of about 10-20\% in $\sim$10-20 days, but the data were not conclusive. Faster variations of the $^{28}$SiO $v$=1,2 $J$=1--0 lines were tentatively reported by \cite{balister77} in observations spaced 4 days apart towards several types of variable stars (Mira-type, SRa, and SRc variables), but the relatively low sensitivity prevented the authors from making any firm conclusions. On these timescales, the expected variations are small and their detection requires spectra with a high signal-to-noise ratio (S/N) and very careful calibration.

In this paper, we present the detection of fast SiO line variability on short timescales in observations spaced a few days apart. We discuss the properties and the physical origin of the fast variability. This behavior is mainly observed in the semi-regular variable RX\,Bootis, but other variable stars are studied to compare the results. To obtain accurate results,  particular care was taken with the relative calibration of the data, as we discuss in Sect.\,\ref{calibration}. We also focus on the possible effects of linear and circular polarization in our SiO maser observations. SiO masers often show a degree of linear polarization of $\sim$30\%, but this can reach almost 100\% in some spectral components \citep{herpin06}. These values are lower in semi-regular variables. Circular polarization is supposed to be lower than 10\% \citep{kemball97,herpin06}. We confirm that our results do not depend on the degree of circular or linear polarization of the SiO lines, and are not due to changes in the polarization degree.

The observations and data reduction are explained in Sect.\,\ref{observations}, together with the absolute and relative calibration, including the careful flux density calibration procedures. The observational results are shown in Sect.\,\ref{results}. We discuss and summarize our conclusions in Sects.\,\ref{sec.discussion} and \ref{conclusions}.

\section{Observations and data reduction}\label{observations}

The studied objects in this work are the prototypical Mira-type stars U\,Her, R\,Leo, and R\,LMi, the semi-regular variables RX\,Boo and RT\,Vir, and the S-type Mira variable $\chi$\,Cyg (See Table \ref{tab.sourcesandobservations}). We carried out observations with a minimum spacing of one day, searching for fast variations in the intensity of the SiO maser lines. We observed the 7, 3, and 2\,mm ($J$=1--0, $J$=2--1 and $J$=3--2) SiO lines. Our observations were performed using the 40\,m antenna at Yebes Observatory (Guadalajara, Spain) and the IRAM 30\,m at Pico Veleta (Granada, Spain). Both telescopes were used simultaneously during the August--September 2017 observing run. 

Calibration of the observations was carefully performed. Poor scans were rejected during the data reduction, whereas the good ones were averaged for each observed day. In order to improve the calibration, we observed thermal lines with high signal-to-noise ratio (S/N) when possible (see Sect.\,\ref{calibration}).
Our observations required a total telescope time of $\sim$ 450 hours distributed over six observational runs. As a result, we obtained a total of about 600 average spectra.

\begin{figure}[h]
   \centering{\resizebox{7.5cm}{!}{
   \includegraphics{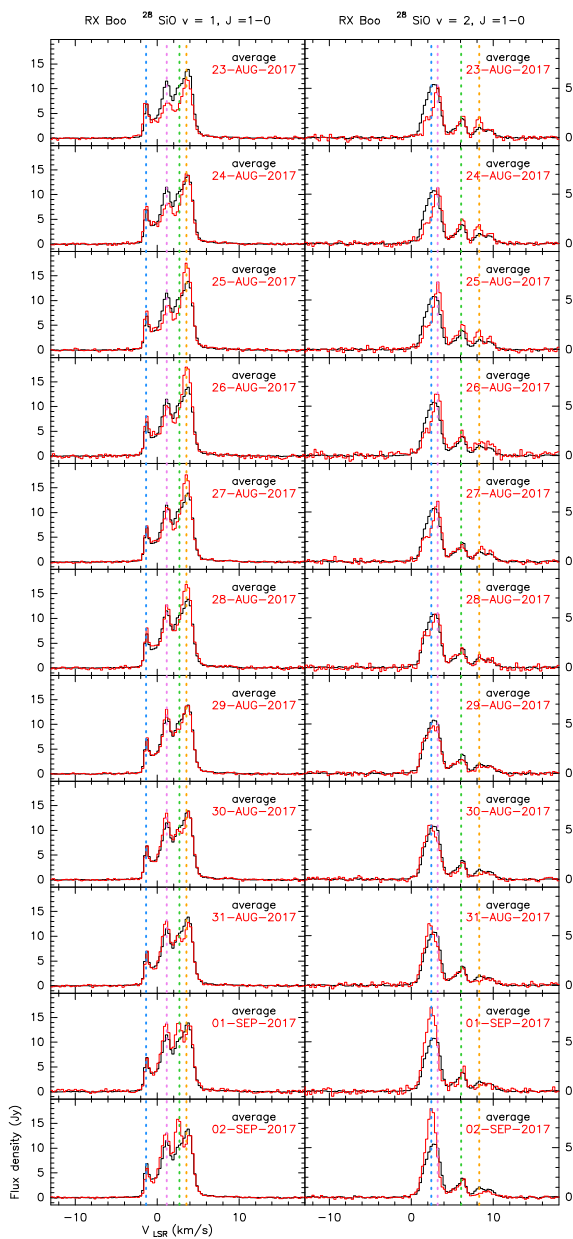}}}
   \caption{Time evolution of the $^{28}$SiO $v$=1 $J$=1--0 (\textit{left}) and $^{28}$SiO $v$=2 $J$=1--0 (\textit{right}) lines observed in RX\,Boo during our August--September 2017 observing run. In both cases, the average spectrum of the complete run is plotted in black and the individual (one per day) spectra are plotted in red. The dashed vertical lines indicate the spectral components, selected by eye (Sect.\,\ref{results.rxboo}), for which our statistical analysis is performed.}
    \label{rxboo_aug2017_V1V2_RL_run}%
\end{figure}

   \begin{figure}[h]
   \centering{\resizebox{8.0cm}{!}{
   \includegraphics{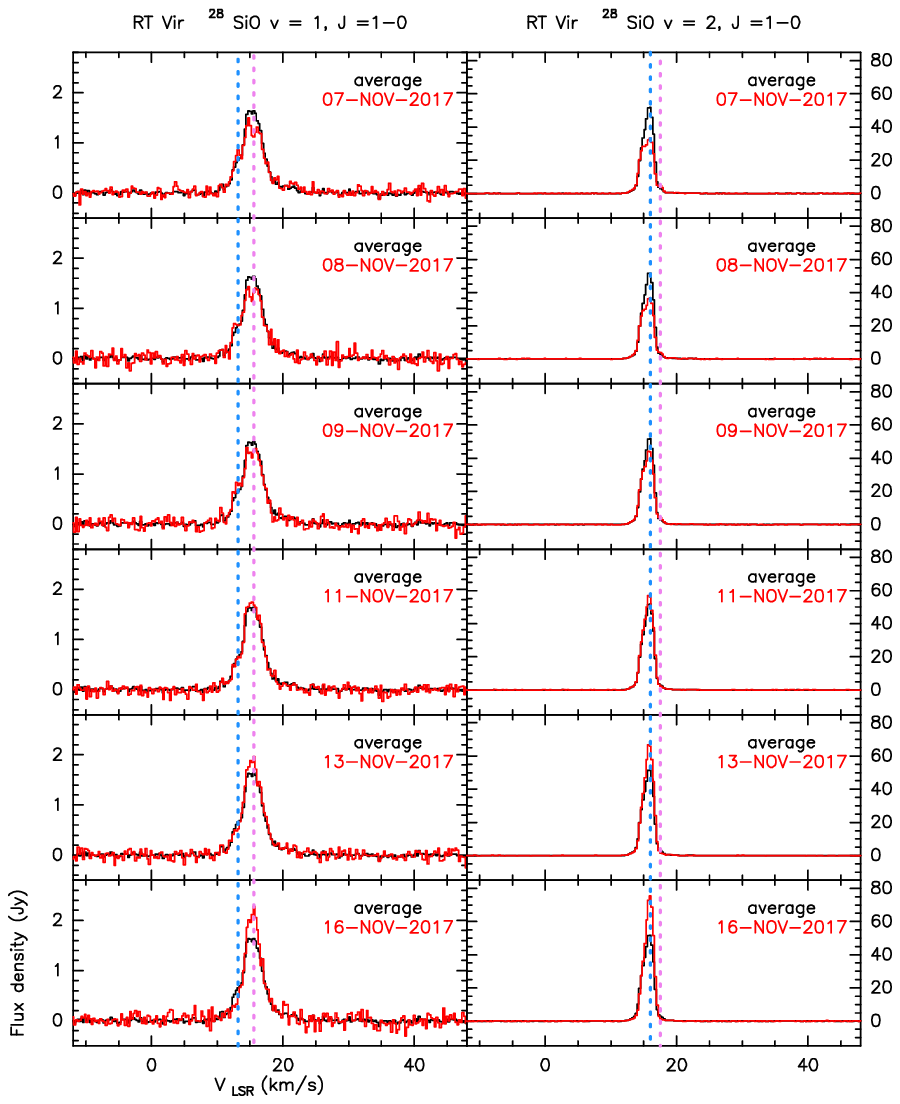}}}
   \caption{Same as in Fig.\,\ref{rxboo_aug2017_V1V2_RL_run}, but for the observations of RT\,Vir in November 2017.}
    \label{rtvir_nov2017_v1_1-0_all}%
\end{figure}

\begin{figure}[h!]
   \centering{\resizebox{7.1cm}{!}{
   \includegraphics{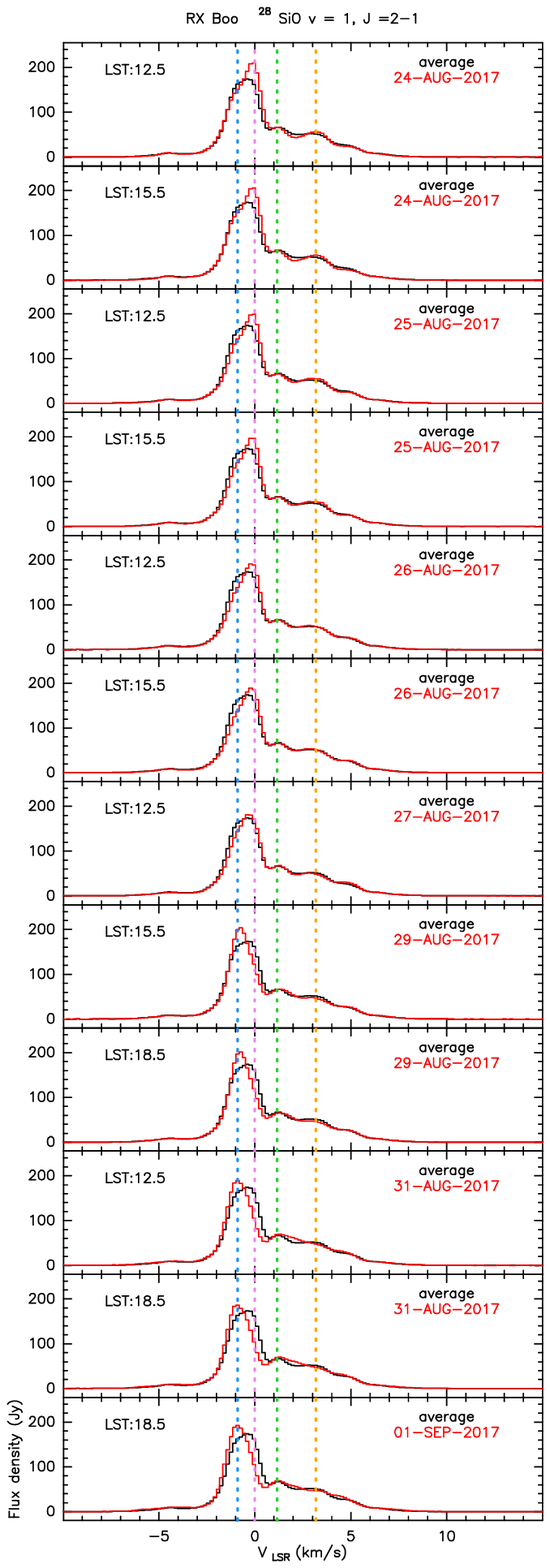}}}
   \caption{Time evolution of the $^{28}$SiO $v$=1 $J$=2--1 line observed in RX\,Boo during our August--September 2017 observing run. The average spectra are plotted in black and the individual spectra (one per local sidereal time and day) are plotted in red. The dashed vertical lines indicate the spectral components for which our statistical analysis is performed.}
    \label{rxboo_aug2017_v1_2-1_all}%
\end{figure}

\subsection{Yebes 40\,m: $\lambda$ = 7\,mm} \label{obsYebes}

We observed the 7\,mm SiO lines using the Yebes 40\,m telescope. Our observations were performed during five different runs: August 2016, September 2016, March 2017, August--September 2017, and November 2017 (see Table\,\ref{tab.sourcesandobservations}). $^{28}$SiO ($v$=0,1,2,3 $J$=1--0) lines and transitions of its isotopic species $^{29}$SiO ($v$=0,1 $J$=1--0) and $^{30}$SiO ($v$=0,1 $J$=1--0) were observed. The received signal was detected by \textit{Fast Fourier Transform Spectrometer} (FFTS) backend units, with an instantaneous bandwidth of 2.5 GHz, and a spectral resolution of $\sim$0.27 kms$^{-1}$ at 43 GHz. The half power beam width (HPBW) of the 40\,m at this frequency is $\sim$43 arcsec; see \cite{devicente16} for more technical details of this system.

The sources were observed using the position-switching method, which was applied to the source and a reference position, with a separation of $\sim$200\,arcsec in the azimuth direction, obtaining flat baselines. Weather conditions were good in all observational runs with a precipitable water vapor of around 4 and 12\,mm in winter and summer epochs,  respectively. Focus and pointing were corrected previous to the integration on each source. When the source exhibited strong maser emission, usually in vibrational states $v$=1 or $v$=2, these lines were used to correct the pointing and focus parameters of the telescope using pseudo-continuum mode on the steady source \citep[see][]{devicente16}. Otherwise, we used a nearby source with intense SiO maser emission for that purpose. Typically, the distance between the source and the pointing calibrator was smaller than 30 degrees. In addition, the flux density calibration was improved using the five-point method \citep{alcolea99} during the observation runs of March and August--September 2017. This data acquisition method provides five spectra: nominal position and four offsets in azimuth and elevation with a typical distance of 12\,arcsec. During the data-reduction phase, the intensities of the SiO maser lines in the five positions were fitted to a Gaussian distribution in order to obtain a more accurate measurement of the pointing errors. In this case, the average spectrum is multiplied by a factor to correct the new pointing errors. The pointing errors were smaller than $\sim$10\,arcsec, which yields flux density corrections smaller than 10\%.

Including pointing and focus calibrations, RX\,Boo and RT\,Vir were observed for 1.5 hr each day. All the scans were averaged yielding one individual spectrum per day per source. Taking into account the fact that the rms obtained around the SiO lines was about 15\,mK, the maser lines were detected with a S/N greater than 10 in RX\,Boo and RT\,Vir. The telescope times devoted to U\,Her, R\,Leo, R\,LMi, and $\chi$\,Cyg were 1.5\,hr each per day. The rms in the daily averaged spectra was about 20\,mK, and the SiO maser lines in these stars were detected with a S/N of greater than 30.

The observations were carried out with dual circular polarization (right and left). With the purpose of evaluating the influence of circular polarization in the detected variations, we analyzed right and left polarizations separately. The evolution of SiO maser lines in right and left circular polarizations matched in time, and the spectral features varied simultaneously. The ratio between LCP and RCP remained approximately constant, with small differences of $\sim$15\% or less, perhaps due to calibration uncertainties (See Fig.\,\ref{rxboo_aug2016_V1V2_RL_comp}). The time evolution of both polarizations was checked for all runs and sources. The circular polarization degree was found to be lower than 5\,\%, which is lower than the relative calibration uncertainties and is often undetectable. As a result, the variations in the circular polarization degree cannot significantly affect the measured fast variability. The right and left circular polarizations were averaged, thereby improving the S/N. Our observations are shown in Figs.\,\ref{rxboo_aug2017_V1V2_RL_run} and \ref{rtvir_nov2017_v1_1-0_all}, and Appendices\,\ref{appenSR} and \ref{appenMiras}. 

\subsection{IRAM 30\,m: $\lambda$ = 2 and 3\,mm}

We carried out observations at 2 and 3\,mm with the IRAM 30\,m telescope at Pico Veleta in August--September 2017. The observations were performed under typical summer conditions with a precipitable water vapor of $\sim$8\,mm. Only scans obtained under stable atmospheric conditions were considered. In this run we observed RX Boo, U Her, and $\chi$\,Cyg (see Table\,\ref{tab.sourcesandobservations}). 

With the purpose of having high spectral resolution, we connected \textit{fast fourier transform spectrometer} (FTS) units  to the EMIR receiver with 50 kHz per channel. The $^{28}$SiO $v$=0,1,2 $J$=2--1 lines and the $^{29}$SiO $v$=0, $J$=2--1 line (around 86 GHz) were observed with a spectral resolution of $\sim$0.17 kms$^{-1}$. The $^{28}$SiO $v$=1,2 $J$=3--2 lines and the $^{29}$SiO $v$=0 $J$=3--2 line at 2\,mm (129 GHz) were observed with higher resolution $\sim$0.11 kms$^{-1}$. To increase the S/N, the spectral resolution was degraded to $\sim$0.22 kms$^{-1}$. Thus, the spectral resolutions at 3 and 2\,mm are comparable.  

The HPBW of the 30\,m telescope is 29 and 19 arcsec at 86 and 129\,GHz, respectively. The sources were observed using the wobbler-switching mode, which provided flat baselines. The subreflector was wobbled every 2 seconds with a throw of 120\,arcsec in the azimuthal direction.

We obtained spectra for the vertical and horizontal linear polarizations. The projections on the sky of vertical and horizontal polarizations change when the parallactic angle varies during the observations. Those variations can be significant if the source is polarized. To minimize that dependence, the observations were performed in parallactic angle ranges smaller than 20 degrees. The selected parallactic angle ranges for RX Boo were: 44$\pm$10, 308$\pm$10, and 61.6$\pm$0.2 degrees, corresponding to three different sidereal time ranges. $\chi$ Cyg was observed in a parallactic angle of 296$\pm$5 degrees. U Her was observed in two different ranges: 314$\pm$10 and 2$\pm$8 degrees. While averaging the horizontal and vertical linear polarizations over a parallactic angle range, differences of $\sim$15\% in intensity between both linear polarizations were observed (see Appendix \ref{appenMaserPol}). However, the time evolution of the vertical and horizontal linear polarizations is simultaneous and the same variations are found in every case (see Fig.\,\ref{rxboo_aug2017_V1_HV_comp}). Again, we conclude that the observed variations of the total flux density cannot be due to variations in the linear polarization degree. As a result, both linear polarizations, that is horizontal and vertical, are averaged at 3\,mm. 

Due to technical problems, the 2\,mm observations between 23 and 27 August 2017 were only performed in vertical polarization resulting in a lower S/N. As a result, the collected information at 2\,mm did not allow us to make any conclusions about the fast variability at this frequency and is not discussed further in this paper. Our observations are shown in Fig.\,\ref{rxboo_aug2017_v1_2-1_all} and Appendix\,\ref{appenMiras}.

  \begin{figure}[h]
   \centering{\resizebox{8cm}{!}{
   \includegraphics{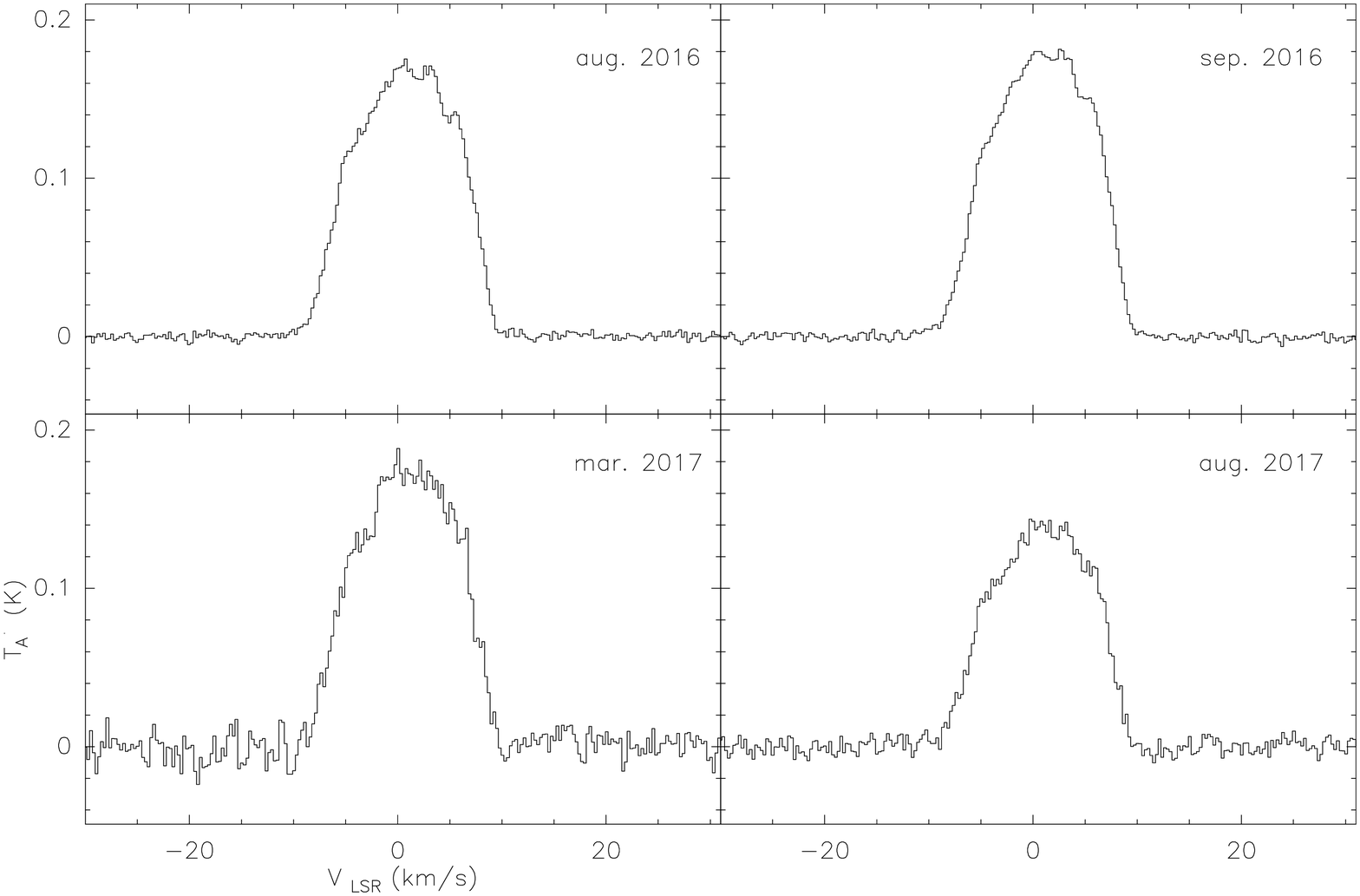}}}
   \caption{$^{28}$SiO $v$=0 $J$=1--0 profiles in the semi-regular variable RX\,Boo for all runs after standard calibration, but before relative calibration. The units and dates of the observations are indicated.}
    \label{rxboo_v0_4runs}%
    \end{figure}

\subsection{Relative flux density calibration}\label{calibration}
At 7\,mm (Yebes 40\,m), the calibration was performed comparing with a hot load and the sky. The calibration at 3\,mm (IRAM 30\,m) was derived using the chopper-wheel method. The procedure was regularly repeated (15--20 min) depending on the weather conditions. Comparing the intensity of constant lines observed in CW\,Leo and NGC\,7027, we estimate the absolute calibration uncertainty to be about 15\% at 3\,mm. At 7\,mm, we obtained an absolute calibration uncertainty of between 15 and 20\% using the intensity of the $^{28}$SiO $v$=0 line in RX\,Boo.

In order to minimize the calibration uncertainty, the spectra were relatively calibrated using $^{28}$SiO $v$=0 emission. The $^{28}$SiO $v$=0 ($J$=1--0 and $J$=2--1) lines are moderately intense in all sources of our sample. These lines are mostly produced by thermal emission, which is not expected to show important changes in intensity. Some $v$=0 lines show narrow spectral features, which are typical features of weak maser emission, overlapping with the thermal profile \citep{devicente16,gonzalez03}. In these objects, the relative calibration is only made using the part of the spectrum without maser features. In a few cases, it was impossible to carry out this procedure because the maser features were present across the whole profile.

We applied secondary corrections ($\sigma_{sec}$) smaller than 10\% at 7 and 3\,mm. Exceptionally, a secondary correction of about 15\% was applied in RX\,Boo at 3\,mm because  of poor weather conditions on 29 August. Taking into account the measured rms (Sect.\,\ref{obsYebes}) and the secondary corrections applied to each individual spectrum, the flux density errors were estimated as $\sqrt{\left(F_{\nu}\cdot\sigma_{sec}\right)^2+rms^2}$.

 The calibration of the data in flux density units is explained in detail in Sects.\,\ref{cal7mm} and \ref{cal3mm} for each frequency. We reiterate the fact that, although the absolute and relative calibrations are robust, the variations of the maser lines are more reliable when the changes are differential for the various maser features.

\subsubsection{$\lambda$: 7\,mm}\label{cal7mm}

As our targets can be considered as point sources, the intensity of the spectra is expressed in flux density units. The conversion factor applied from antenna temperature to flux density depends on the aperture efficiency, $\eta_A$, as 2.2\,·\,$\eta_F/\eta_A$, where $\eta_F$ is the coupling factor between the receivers and the sky. We use a value of 0.9 for $\eta_F$, which is derived by fitting skydips. The aperture efficiency  was obtained using continuum observations of sources with well-known flux density and size, such as planets, and was found to be slightly variable from run
to run. Therefore, our flux density scale is directly deduced from observations of such standards and is slightly dependent  on the efficiency uncertainties.

Using accurate observations in RX\,Boo, we derive the flux density of the $^{28}$SiO $v$=0 $J$=1--0 line. Figure\,\ref{rxboo_v0_4runs} shows the $^{28}$SiO $v$=0 $J$=1--0 line in RX\,Boo for all runs before the flux and relative calibration. We find the intensity of this line to be 1.35$\pm$0.15 Jy. This measurement was used to improve the calibration in flux density units at 7\,mm.

The maser lines in the semi-regular variables RX\,Boo and RT\,Vir were relatively calibrated using the $^{28}$SiO $v$=0 $J$=1--0 line, degrading the spectral resolution to 0.5 and 1.3 kms$^{-1}$ respectively. Previous observations in 2014 \citep{devicente16} show that the shape and flux density of these lines are constant over timescales of years.

The data of the S-type star $\chi$\,Cyg were also recalibrated using its $^{28}$SiO $v$=0 $J$=1--0 line. This line shows a maser component superimposed on the thermal emission at the central velocity. The profile is smooth at other velocities. We assume that the emission at these velocities is thermal and can be used for calibration. In order to avoid maser features, a velocity range between 0 and 6\,kms$^{-1}$ was selected. As shown by \cite{devicente16}, the intensity in this velocity range is constant in time.

The S/N of the SiO $v$=0 $J$=1--0 lines is too small or shows maser features in the complete profile of the observed O-rich Mira-type sources. Therefore, U\,Her, R\,Leo, and R\,LMi are not relatively calibrated using thermal emission.

\subsubsection{$\lambda$: 3\,mm}\label{cal3mm}

Although the HPBW is somewhat smaller in the observations with the 30\,m IRAM telescope, the observed objects can also be considered to be point sources at these frequencies. The conversion factor used was $\sim$6 Jy/K, derived from the IRAM handbooks.

The $^{28}$SiO $v$=0 $J$=2--1 line presents in\,RX Boo and $\chi$\,Cyg a typical thermal line profile. Previous observations show that these lines are constant in time \citep{lucas92,bujarrabal91}. We checked that the applied calibration was consistent with the intensity of the $^{29}$SiO $v$=0 $J$=2--1 line, which is also assumed to be thermal. 

The secondary calibration of U Her data was harder at 3\,mm, since the SiO $v$=0 line shows maser features at negative velocities. Taking into account this constraint, a velocity range from -9.7 to 2.8\,kms$^{-1}$ was selected in order to avoid maser emission.

\section{Results}\label{results}
\subsection{RX Boo}\label{results.rxboo}

RX\,Boo is a semi-regular variable and presents the strongest variations on short timescales detected in this work. RX Boo was observed on several dates corresponding to different phases of its pulsation (August 2016: 0.9; September 2016: 0.1; March 2017: 0.5, August-September 2017: 0.6). The phases were estimated using the available observations from \textit{AAVSO} (https://www.aavso.org) and assuming a period of $\sim$160\,days taken from the \textit{General Catalog of variable stars} \citep{samus17}. 

Figures\,\ref{rxboo_aug2017_V1V2_RL_run}, \ref{rxboo_aug2016_V1V2_RL_run}, \ref{rxboo_sep2016_V1V2_RL_run}, and \ref{rxboo_mar2017_V1V2_RL_run} show the resulting spectra obtained during the observations performed with the 40\,m Yebes radio telescope. $^{28}$SiO $v$=1 and $v$=2 are plotted for each observational run. The average spectrum of the complete run is plotted in black to ease the comparison of the time evolution. The individual (one per day) spectra are plotted in red.
   
From the data of the four observational runs, we conclude that strong and fast variations are detected in our observing runs. Such variations do not depend on the vibrationally excited state. For a particular observational run, the fast variability can appear at $v$=1, $v$=2, or both. We also verified that there are no velocities where variations are preferentially seen, and that the variability can appear in any velocity. In the particular case of the observations performed during August 2017, our most complete observing run, the masers in the vibrationally excited states $v$=1 and $v$=2 show strong changes (see Fig.\,\ref{rxboo_aug2017_V1V2_RL_run}). In this run, both vibrationally excited states, $v$=1 and $v$=2, show strong variations with a similar time evolution for the spectral feature at $\sim$2.5 kms$^{-1}$. This suggests that the emissions of both transitions at this velocity come from the same spot.
   
Although the variations of the maser components are occasionally significant between observations separated by one day, the most remarkable changes are often seen in observations separated by two or three days. Comparing the observations spaced by two days, variations of greater than 50\% in the intensity are clearly seen. These strong and fast variations have not been previously reported. 

We detected weak $^{28}$SiO $v$=3 $J$=1--0 maser emission showing fast variability, but  this could not be properly studied because of the poor S/N ratio. The $J$=1--0 emission in the vibrational state $v$=0 of the isotopic species $^{29}$SiO and $^{30}$SiO was detected with low intensity. As it was impossible to properly analyze those profiles in the individual spectra, we averaged the spectra for each run. These lines do not seem to present maser features and the profiles are similar to that of the $^{28}$SiO $v$=0 line. We therefore conclude that these lines are formed by thermal emission and they probably show constant intensity over time. Other vibrational states of $^{29}$SiO and $^{30}$SiO are not detected in RX Boo.

We also observed transitions of SiO and its isotopes towards RX\,Boo at 3\,mm in August 2017. The $^{28}$SiO $v$=1 $J$=2--1 maser line was detected with an intensity $>$ 200 Jy. We reached a high S/N allowing an accurate analysis. The $^{28}$SiO and $^{29}$SiO $v$=0 $J$=2--1 lines were detected, but they do not exhibit maser emission. Figure\,\ref{rxboo_aug2017_v1_2-1_all} shows the sequence of $^{28}$SiO $v$=1 $J$=2--1 spectra for several dates and sidereal times. Local sidereal time is quoted to accurately indicate the time lapsed between each observation. The average spectrum is also plotted to highlight the observed variations of the individual spectra. 

\begin{figure}[h!]
   \centering{\resizebox{8.0cm}{!}{
   \includegraphics{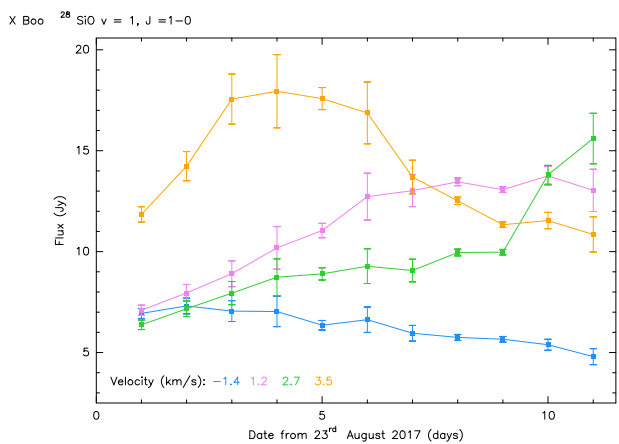}}}
   \caption{Time evolution of the $^{28}$SiO $v$=1 $J$=1--0 spectral components observed in RX\,Boo during our August--September 2017 observing run. The different colors indicate the velocity of each spectral feature. }
    \label{rxboo_aug2017_v1_1-0_var}%
\end{figure}

\begin{figure}[h!]
   \centering{\resizebox{8.0cm}{!}{
   \includegraphics{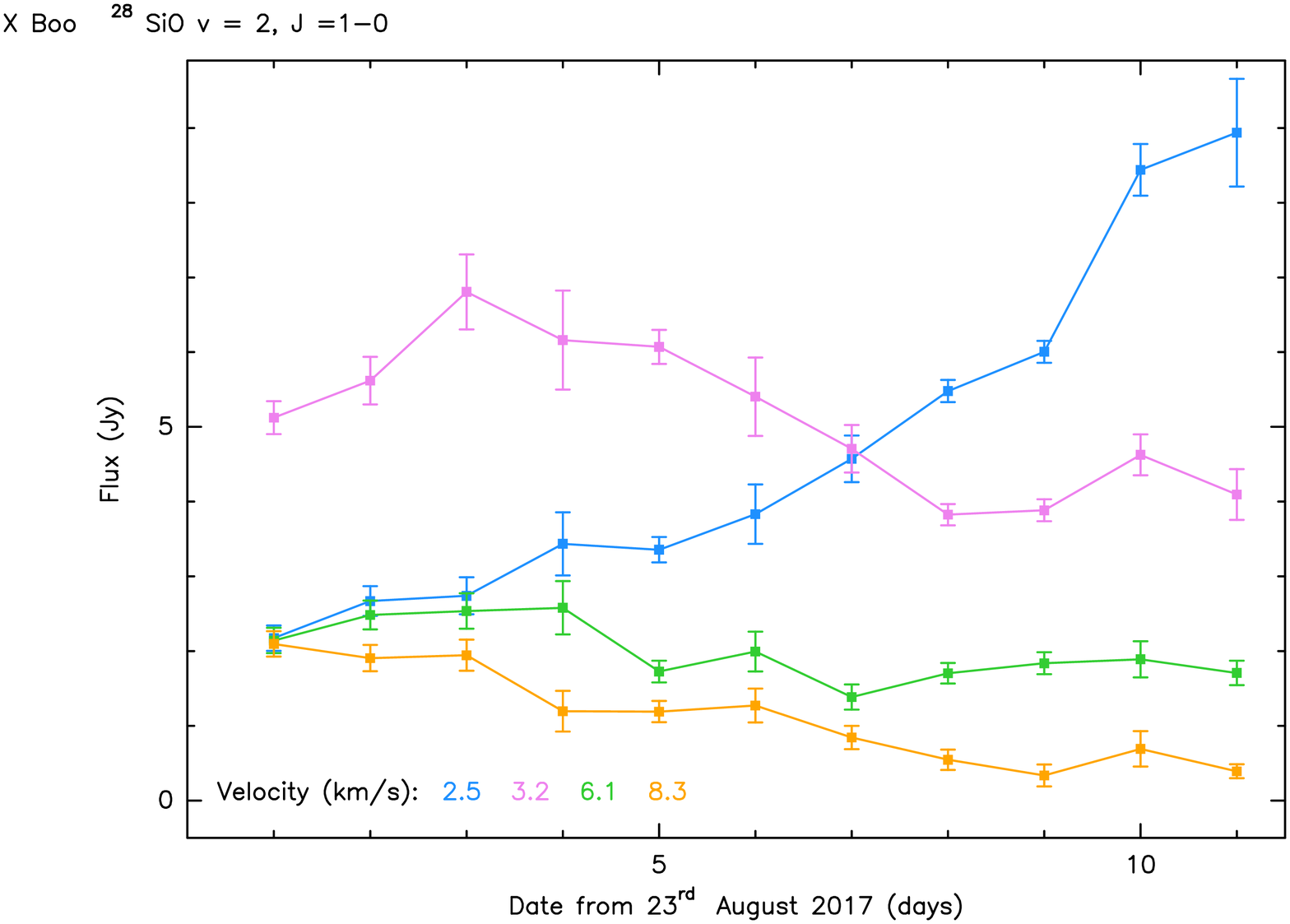}}}
   \caption{Time evolution of the $^{28}$SiO $v$=2 $J$=1--0 spectral components observed in RX\,Boo during our August--September 2017 observing run. The
different colors indicate the velocity of each spectral feature. }
    \label{rxboo_aug2017_v2_1-0_var}%
\end{figure}

\begin{figure}[h!]
   \centering{\resizebox{8.0cm}{!}{
   \includegraphics{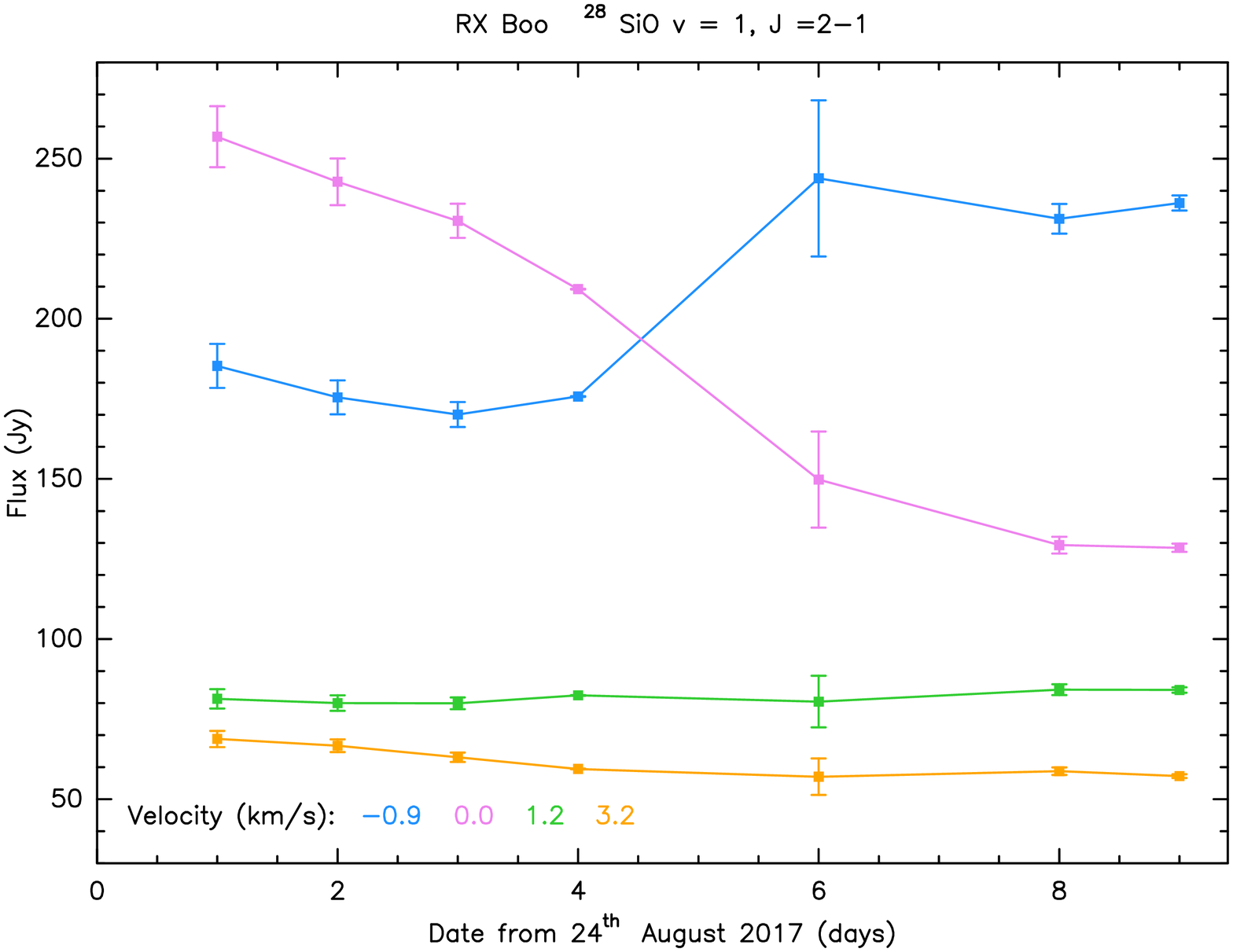}}}
   \caption{Time evolution of the $^{28}$SiO $v$=1 $J$=2--1 spectral components observed in RX\,Boo during our August--September 2017 observing run. The
different colors indicate the velocity of each spectral feature. }
    \label{rxboo_aug2017_v1_2-1_var}%
\end{figure}

A change in the velocity of the line peak is clearly seen over a few days. We note the abrupt variation detected between August 27 and 29, where the velocity of the emission peak varies by a total of $\sim$0.6 kms$^{-1}$. This change in the velocity could be the result of the appearance of a new spectral feature in the spectra, which would indicate that the emission is coming from two independent emitting regions. On the other hand, it is also possible that there was an abrupt variation in the velocity of a single emitting clump. However, the available information is not enough to discern the origin of this particular behavior. 

After inspection of the daily and averaged spectra by eye, we identified several velocities that correspond to secondary maxima or bumps and were judged to represent the whole profile shape; these are shown by dashed vertical lines with different colors in the spectra. To quantify the variation rate of the maser lines, we plot the intensity at the selected velocities for each individual observation. The results can be seen in Figs.\,\ref{rxboo_aug2017_v1_1-0_var}--\ref{rxboo_aug2017_v1_2-1_var} and Appendix\,\ref{appenSR}. The variations in the intensity of a single spectral feature are often independent of the variations in the other maser components, confirming that our general results are indeed independent of the relative flux density calibration.

To quantify the variation rate, we performed a statistical analysis (see Appendix \,\ref{appenStat}). Figure \,\ref{histograms_rxboo_aug2017_V1} presents an example of our analysis for the $^{28}$SiO $v$=1 $J$=1--0 observations in August--September 2017, which is our best sampled observing run. The histograms show the number of times that the variation of a single maser feature is higher than 10\%, 20\%, 40\%, and 60\% for the total set of "experiments" performed. We define an "experiment" in this context as a measurement of the intensity variation obtained for any maser feature and any time spacing (we consider 1, 2, 4, and 7 days). For a given experiment, the relative variation is defined as the largest measured difference between intensities within the considered time period divided by the average intensity  during that period. A summary of the results can be seen in Appendix\,\ref{appenStat} for the particular cases of 10\% and 40\% variations of the intensity (see Tables \ref{tab.var10-40V1aug16_1-0}--\ref{tab.var10-40V2aug17_1-0}).

\begin{figure*}[h!]
   \centering{\resizebox{18.cm}{!}{
   \includegraphics{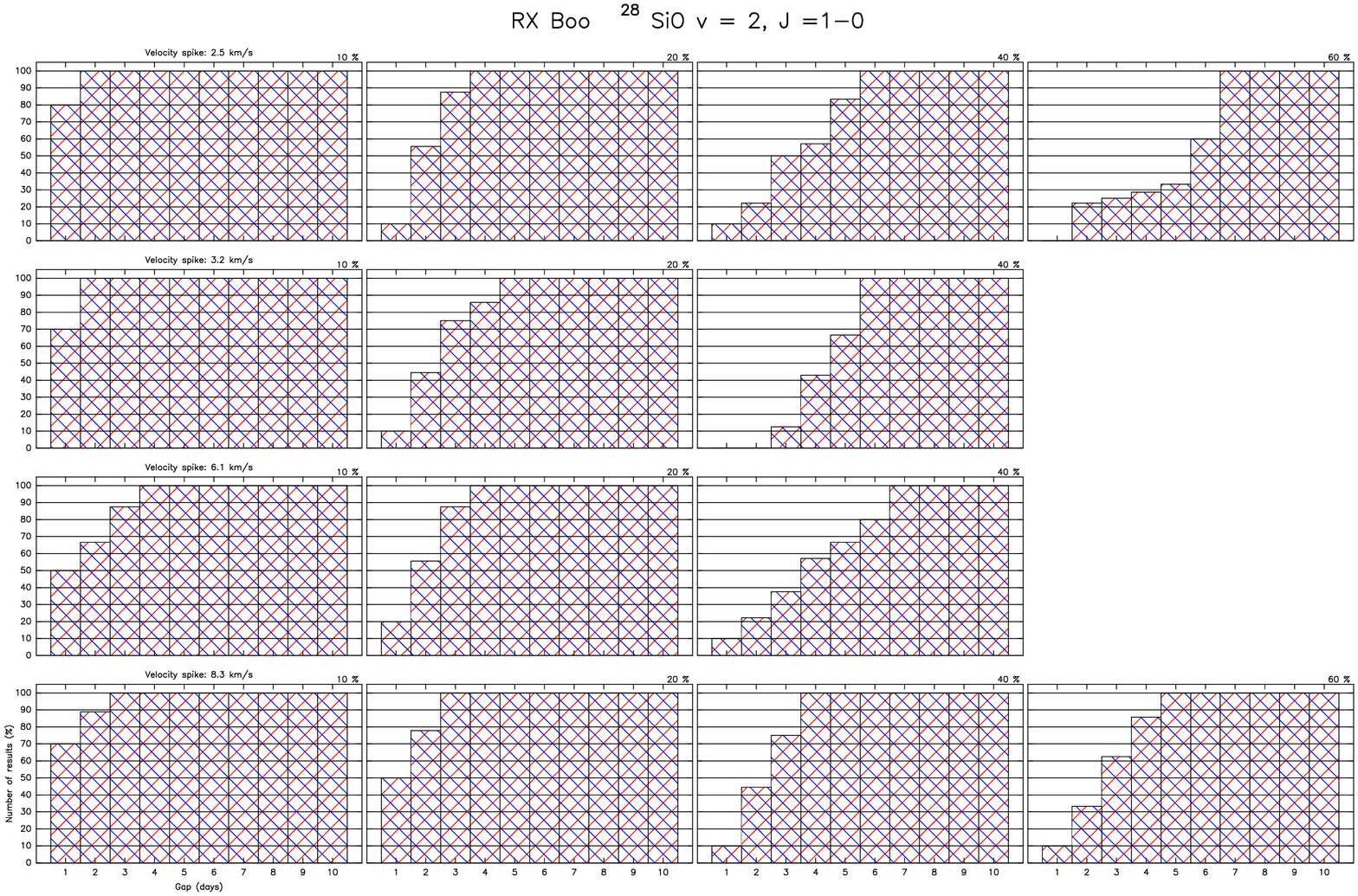}}}
   \caption{Statistical analysis of the $^{28}$SiO $v$=2 $J$=1--0 line observed in RX\,Boo during August--September 2017 for variation rates of 10\%, 20\%, 40\%, and 60\%. We conducted this analysis for four spectral features: 2.5, 3.2, 6.1, and 8.3 kms$^{-1}$}
    \label{histograms_rxboo_aug2017_V1}%
   \end{figure*}

\begin{figure}[h]
   \centering{\resizebox{8.0cm}{!}{
   \includegraphics{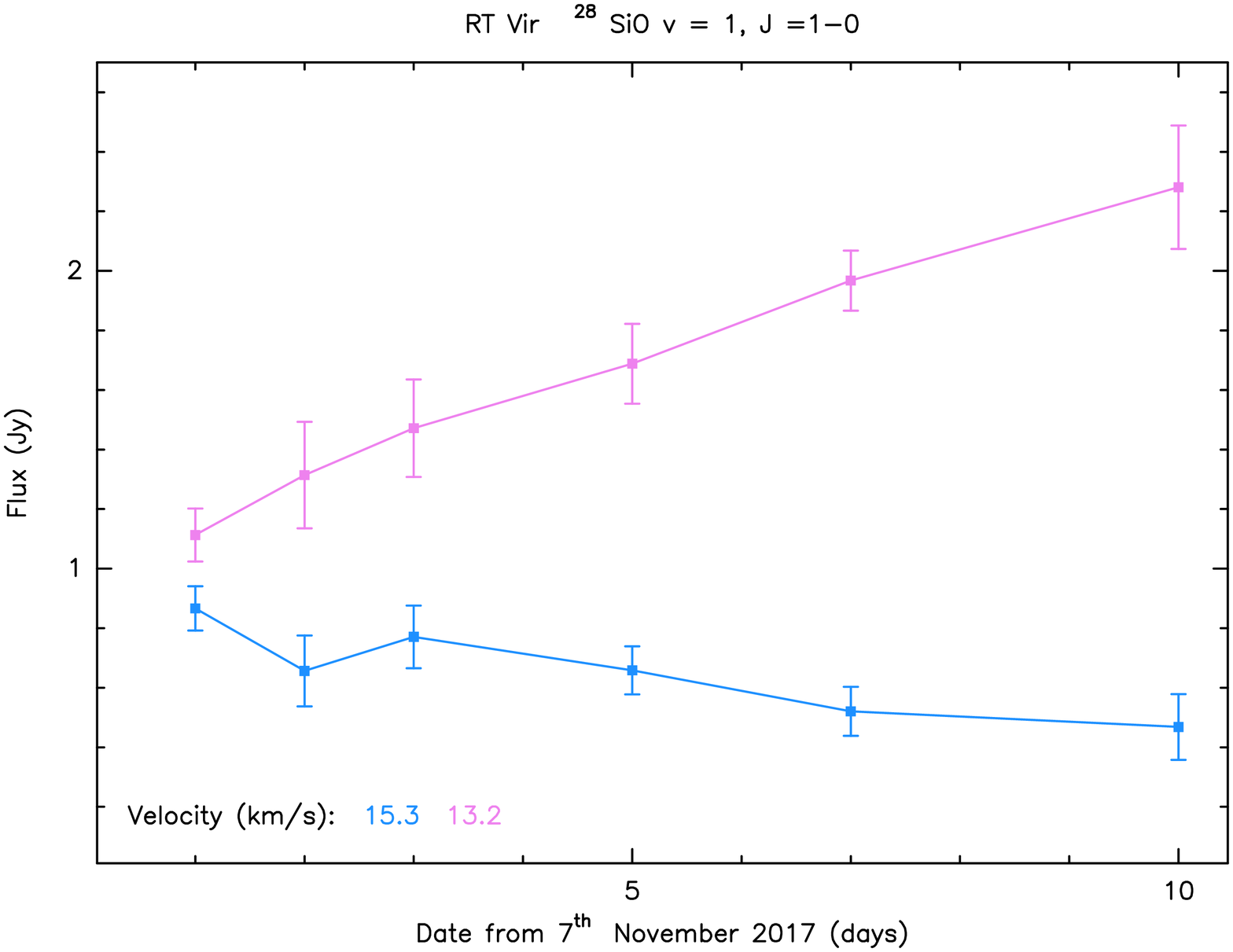}}}
   \caption{Same as in Fig.\,\ref{rxboo_aug2017_v1_1-0_var}, but for the observations of RT\,Vir in November 2017.}
    \label{rtvir_nov2017_v1_1-0_var}%
\end{figure}

\begin{figure}[h]
   \centering{\resizebox{8.0cm}{!}{
   \includegraphics{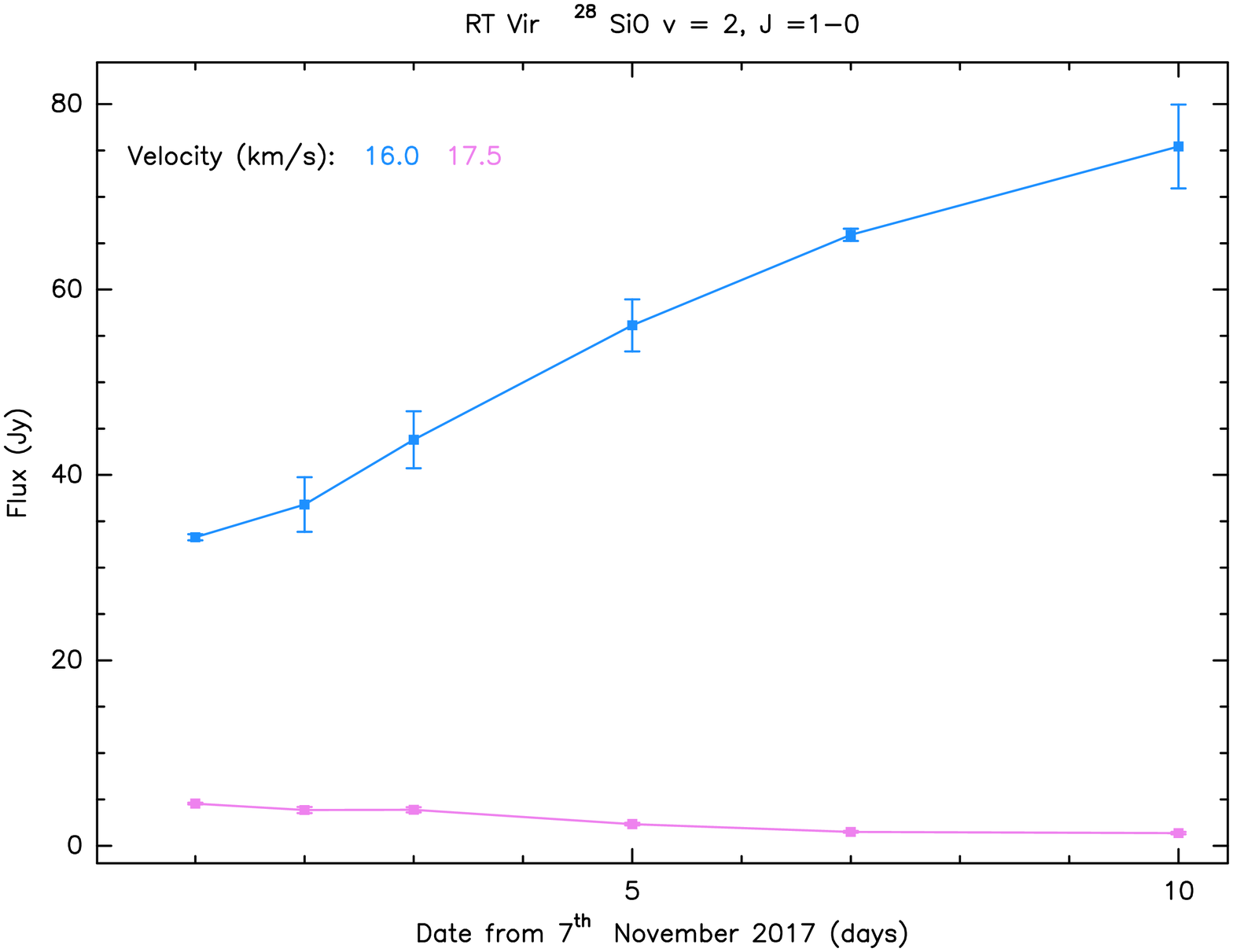}}}
   \caption{Same as in Fig.\,\ref{rxboo_aug2017_v2_1-0_var}, but for the observations of RT\,Vir in November 2017.}
    \label{rtvir_nov2017_v2_1-0_var}%
\end{figure}
    
\subsection{RT Vir}\label{results.rtvir}

The other semi-regular variable of our sample, RT\,Vir, was only observed in one run (see Table\,\ref{tab.sourcesandobservations}) and at 7\,mm. Figure\,\ref{rtvir_nov2017_v1_1-0_all} shows the time evolution of the $^{28}$SiO $v$=1,2 $J$=1--0 maser lines in RT Vir. The profile exhibits strong variations on scales of a few days, although they are somewhat weaker than for RX\,Boo. In particular, the $^{28}$SiO $v$=2 $J$=1--0 line can vary by $\sim$50\% in 4\,days. Figures\,\ref{rtvir_nov2017_v1_1-0_var} and \ref{rtvir_nov2017_v2_1-0_var} show the quantitative time evolution of the intensity of two spectral features identified as bumps or secondary maximums upon inspection. As for RX\,Boo, a relative change between the intensities at different velocities is found. Following the same procedure described for RX Boo, we perform a statistical analysis of the RT\,Vir data. Tables\,\ref{tab.rtvir_var10-40V1nov17_1-0} and \ref{tab.rtvir_var10-40V2nov17_1-0} are a summary of the results for the particular cases of 10\% and 40\% variation rate. Although our observations of RT\,Vir are less complete than those of RX Boo, we find that this source also presents a significant variation over a time period of a few days.
We also detected $^{28}$SiO $v$=3 $J$=1--0; this line shows maser emission and significant variability. The $^{29}$SiO and $^{30}$SiO $v$=0 $J$=1--0 lines were also seen to show apparent thermal emission.

\subsection{Mira-type variables: U Her, R LMi, R Leo, and $\chi$ Cyg.} \label{sect.miras}

Mira-type variable stars exhibit a regular long-period variability curve of the maser lines (Sect.\,\ref{intro}). U\,Her, R\,LMi, R\,Leo, and the S-type $\chi$\,Cyg were observed in March 2017 and August--September 2017 (see Table\,\ref{tab.sourcesandobservations}). The time evolution of the $^{28}$SiO $v$=1,2 $J$=1--0, and $v$=1 $J$=2--1 lines is shown in Appendix\,\ref{appenMiras}. 

Changes in the intensity of the maser features of $\sim$10\% are seen after 7 days for these objects. This agrees with the variation rate expected for the regular variables, and previously observed by \cite{pijpers94} and \cite{pardo04}. Only one spectral component of the $v$=2 $J$=1--0 line in R\,Leo shows stronger variation during the observational run of March 2017. The maximum variation rate observed in this case reaches a value close to 15\% after 2 days. It is important to note that R\,Leo exhibits a SiO maser long-term variability that is slightly less regular and can present erratic behavior, as was shown by \cite{pardo04}. Faster but much smaller ($\sim$\,2\% ) and roughly periodical variations of the linear and circular polarization degree were observed by \cite{wiesemeyer09} in this source, but such a phenomenon does not seem related to the total flux density variations found here.

We also detected $^{28}$SiO $v$=3 $J$=1--0, $^{29}$SiO $v$=0 $J$=1--0, $^{30}$SiO $v$=0 $J$=1--0, and $^{29}$SiO $v$=0 $J$=2--1. All of these lines seem to be masers (see Sect.\,\ref{intro}), except for $^{29}$SiO $v$=0 $J$=2--1 in $\chi$\,Cyg. No significant variations were detected, which in some cases is due to the low S/N. The $^{28}$SiO $v$=1 variations found in $\chi$\,Cyg in August--September 2017 (Fig.\,\ref{chicyg_aug2017_v1_1-0_all}) are also unreliable because of the low S/N and the simultaneous variations of all spectral components.

\section{Discussion}\label{sec.discussion}
\subsection{Fast variability of SiO masers in semi-regular variables} \label{sect.disSR}
The long-term monitoring of the SiO masers in the Mira-type variables \citep[see Sect.\,\ref{intro},][]{pardo04} shows that SiO emission is strongly variable and follows the infrared stellar variability, with a period of 200 -- 500 days. However, no significant variations  have been found on timescales shorter than $\sim$ 30 days,  until now \citep[see also][]{pijpers94}. The SiO masers in semi-regular variables, although less well studied, were found to show weaker and more erratic variability on long timescales.

Our observations reveal very fast variability of the SiO maser lines in the semi-regular variable RX\,Boo. In order to summarize our statistical results shown in Tables\,\ref{tab.var10-40V1aug16_1-0} -- \ref{tab.var10-40V1aug17_2-1}, we reiterate that here we define an "experiment" as a measurement of the flux density variation rate between two observations for a given spectral feature. All the spectral features defined in Sect.\,\ref{results} and all possible pairs of observations, for any time gap, are considered. Let us take the August--September 2017 run and the $v$=2 $J$=1--0 line  as an example. Table\,\ref{tab.var10-40V2aug17_1-0} indicates that for the spectral component at 3.2 kms$^{-1}$, 70\% of the experiments for a gap of 1 day show a flux density variation of at least 10\%. For the spectral component at 6.1 kms$^{-1}$, 67\% of the experiments for a gap of 2 days show this variation rate. Variations as high as 40\% in the spectral feature at 2.5 kms$^{-1}$ are observed in 57\% of the experiments for a gap of 4 days.

We can see (Tables\,\ref{tab.var10-40V1aug16_1-0} -- \ref{tab.var10-40V1aug17_2-1}) that, in all our observations of RX\,Boo, there is at least one spectral feature for which a 10\% variation appears in at least one-quarter of the experiments in a gap of only 1 day. For a gap of 2 days, a 10\% variation typically appears  in one-half of the experiments. Variations of 40\% are more rare, but often appear in our data. In the August 2017 run, for the $v$=2 $J$=1--0 transition, about one-half of the experiments for a gap of 4 days show variations of at least 40\%; see Fig.\,\ref{histograms_rxboo_aug2017_V1} for a more complete description of the variability in this run. In March 2017, for $v$=2 $J$=1--0, 67\% of the experiments for a gap of 2 days showed a variation rate of 40\% (Table\,\ref{tab.var10-40V2mar17_1-0}). In the August 2016 run (Table\,\ref{tab.var10-40V1aug16_1-0}), for the $v$=1 $J$=1--0 transition, the same variation rate is measured in about 50\% of the experiments for a gap of 7 days in the spectral component at -0.75 kms$^{-1}$. In September 2016, for $v$=2 $J$=1--0, no experiment reached such a variation rate value even for the longest considered gap of 7 days (Table\,\ref{tab.var10-40V2sep16_1-0}).

The other SRb in our sample, RT\,Vir, although not studied in such detail, also shows fast variability comparable to RX\,Boo (Tables\,\ref{tab.rtvir_var10-40V1nov17_1-0}, and \ref{tab.rtvir_var10-40V2nov17_1-0}). For the analyzed Mira-type variables (U\,Her, R\,Leo, R\,LMi and $\chi$\,Cyg, Sect.\ref{sect.miras}), variations on  timescales of a  few days are almost always under our detection limit (i.e., < 10\%). Only in R\,Leo, during the March 2017 observing run, did one spectral feature show a variation close to 15\% after 2 days (Fig.\,\ref{rleo_mar2017_v1_1-0_all})

\subsection{What is the origin of the observed fast variability?} \label{sect.originFastVar}

We think that the peculiar pulsation mechanism in semi-regular variables, and the resulting series of complex shock fronts traversing the inner circumstellar shells (where SiO maser lines are formed, Sect.\,\ref{intro}) may explain the fast variations of the SiO maser intensity found in RX\,Boo and RT\,Vir. SRb-type variables are mainly overtone pulsators, with shorter optical periods and superposed periodicities, whereas other types of variables (SRa, SRc, and Miras) are characterized by a fundamental pulsation mode \citep{mcintosh15,willson00}. The result is a more chaotic shock structure and a very turbulent environment in the relevant inner shells around SRbs. The structure and dynamics are therefore expected to be very complex, even at relatively small scales, with many small clumps whose physical parameters can change very quickly.

   \begin{figure}[h!]
   \centering{\resizebox{8.5cm}{!}{
   \includegraphics{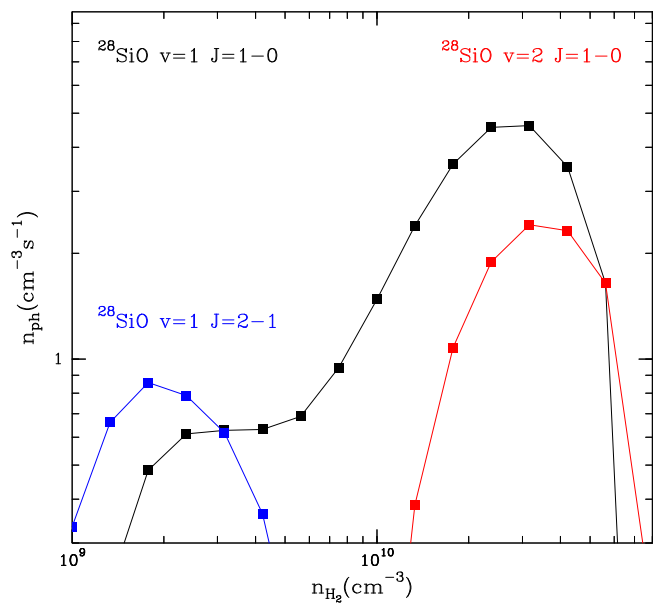}}}
   \caption{Model calculations of the excitation of the $v$=1 and $v$=2 $J$=1--0 SiO maser transitions. The number of maser photons emitted as a function of the gas density is shown.The molecular excitation model includes the effects of the line overlap between the ro-vibrational lines $\nu_2$=0,12$_{7,5}$--$ \nu_2$=1,11$_{6,6}$ of para-H$_2$O and $v$=1,$J$=0--$v$=2,$J$=1 of $^{28}$SiO.}
   \label{fig.modelo}
   \end{figure}

The passage of shocks can significantly affect the properties of the individual emitting clumps, which are responsible for the different maser components observed in SiO lines. The typical velocity of the shocks in AGB inner shells is $\sim\,$10\,kms$^{-1}$ \citep{hinkle78,reid97,hofner03} and therefore they would need at least $\sim$\,60 days to cross a typical maser spot (assumed to be $\sim$\,4--6 10$^{12}$\,cm wide from VLBI observations; Sect.\,\ref{intro}). Therefore, the very fast variations found in RX Boo could be explained either by extremely fast shocks (with velocities $>$ 100\,kms$^{-1}$) or by  the fact that the maser spots are uncommonly small, namely $<$\,10$^{12}$\,cm. The first explanation is unlikely, because such high velocities would induce high projected velocities and high pulsation amplitudes, in contradiction with the observational results (including the moderate SiO line velocities and total luminosity amplitude in SRs) and our general ideas on the movements in the inner circumstellar shells. The second suggestion, on the other hand, seems plausible, but would have important implications on the formation of SiO masers and can only be confirmed by observing at very high angular resolution. As an example, the size of an SiO spots can be estimated using the time evolution of the spectral feature at 3.5 kms$^{-1}$ in the SiO $v$=1 $J$=1--0 line, which was observed in August--September 2017 (Fig.\,\ref{rxboo_aug2017_v1_1-0_var}). If we assume, in this particular case,  that the shock wave passed across the entire SiO spot  in 4 days (to explain the observed increase and decrease of the intensity), we can derive a spot size of about $\sim$4 10$^{11}$cm, which is much smaller than the spots usually found in Mira variables. We note that 4 10$^{11}$cm at the distance of RX Boo, namely 128 pc \citep[taken from the \textit{Gaia} DR2 archive;][]{gaiaColl18}, is equivalent to $\sim$ 0.2 milliarcseconds. In other cases found in our observations, with significant variations within 1 or 2 days, even smaller size scales would be deduced. 

We investigated this second point further, and using a theoretical approach we can estimate the effects of shocks in the maser spots. We performed new calculations of the molecular excitation of the SiO masers following the model proposed by \cite{soriaruiz04}. In this model, the maser flux density is calculated under the $LVG$ approximation, with $d\left(Ln\,V\right)/d \left(Ln\,r\right) =\,\epsilon$\, = 3\,, at a distance of $d$\,=\,2.75\,$R_{\star}$ from the central star. Also, the SiO molecular excitation takes into account the effects of the line overlap between the ro-vibrational lines $\nu_2$\,=\,0,12$_{7,5}$ -- $ \nu_2$\,=\,1,11$_{6,6}$ of para-H$_2$O and $v$\,=\,1,\,$J$\,=\,0 -- $v$\,=\,2,\,$J$\,=\,1 of $^{28}$SiO, which is believed to alter the intensity and spatial distribution of the masers.

 The results are shown in Fig.\,\ref{fig.modelo}, where the number of maser photons as a function of the gas density is plotted. As can be seen, the intensity of the masers is sensitive, although not extremely sensitive, to variations in the density where the masers occur. For instance, changes of $\sim$\,20--30\% in the gas density can induce similar high variations in the maser intensities. This means that the rapid variations that we have found can be explained if the shocks significantly alter the local physical conditions in the envelope, provided that the clump size is small enough, as discussed above.

Further investigation is required to study the effects of shock passage, and of additional phenomena that may be involved in the formation and production of these short-term variations of the SiO masers in SR variables, such as for example the effects of the line overlaps in the maser pumping. A more detailed and systematic VLBI mapping of these emissions in SR variables would be particularly useful.


\section{Conclusions}\label{conclusions}

In order to study the short-term variability of the SiO maser lines in AGB stars, the semi-regular variables RX Boo and RT Vir and the Mira-type variables U Her, R LMi, R Leo, and $\chi$ Cyg were observed in several runs (see Table \ref{tab.sourcesandobservations}). Our observations were carried out using the Yebes 40\,m and the IRAM 30\,m telescopes. SiO maser lines at 7 and 3\,mm, namely the $^{28}$SiO $v$=0,1,2 $J$=1--0 and $^{28}$SiO $v$=1 $J$=2--1 transitions, were observed (see Sect.\,\ref{observations}). In total, we performed five observing runs, each composed of between 8 and 11 individual observations separated by a minimum of 1 day. We discuss the possible effects of the circular and linear polarization (Sect.\,\ref{observations}). The data were very carefully calibrated (Sect.\ref{calibration}), using thermal lines to check the relative calibration when possible, and comparing the various maser components of each observed line. We are convinced that the detected variability is not due to the calibration or variations in the polarization. We stress that the variations of the different spectral features of a given line are in general completely independent. We performed a detailed statistical analysis of the appearance of variations on different timescales, namely of between 1 and 10 days (see Sect.\,\ref{results} and Appendix\,\ref{appenStat}).  Our conclusions can be summarized as follows:

\begin{itemize}

        \item RX\,Boo shows very strong and fast variation in the intensity of the different spectral features of the SiO lines at 7 and 3\,mm (see Sect.\,\ref{results.rxboo}). On a timescale of 1 day, we find variations $\gtrsim$ 10\% in one-quarter of the cases studied (Sect.\,\ref{results.rxboo} and Appendix\,\ref{appenStat}). Variations larger than $\sim$50\% are often found when the observations are separated by 2 or 3 days. This high variation rate has never been detected before in AGB stars, and it could change our understanding of the main properties of the SiO-emitting region, particularly for the semi-regular variables. \\
        
        \item The absolute calibration uncertainty is about 15\% at 3\,mm and between 15 and 20\% at 7\,mm. The relative uncertainty between different observations is smaller than 10\% in any case. The variations found are clearly larger than these uncertainties.\\
        
        \item At 7\,mm, fast variability of the SiO lines in RX\,Boo is seen in the vibrationally excited states $v$=1 and $v$=2, independently , and is detected in every phase of the stellar pulsation. At 3\,mm, the variations of $^{28}$SiO $v$=1 $J$=2--1 lines seem to show a similar behavior to that seen at 7\,mm, but fewer observations at 3\,mm were performed. \\
        
        \item A similar variation rate of the SiO lines at 7\,mm is found in the other semi-regular star in our sample, RT\,Vir, but this source was less well observed (see Sect.\,\ref{results.rtvir}) and the time evolution of the SiO lines at 3\,mm was not investigated.\\
        
        \item The variations of the SiO maser line intensity in the Mira-type variables is very moderate, with variation rates of $\lesssim$10\% after 7 days. Only in R\,Leo  was a change close to 15\% after 2 days found for one of its maser components. \\

\end{itemize}
        
         We suggest that this phenomenon of very fast SiO maser variability only takes place in semi-regular variables. One possible explanation for this phenomenon is the presence of particularly small maser-emitting clumps in such objects, which would lead to a strong dependence of the intensity on the density variations due to the passage of shocks. Further studies are necessary to confirm these general conclusions, including more observations of fast variability and VLBI mapping with very high angular resolution.\\

\begin{acknowledgements}
We are grateful to the anonymous referee for his helpful and constructive comments. This work has been supported by the Spanish Ministry of Science, Innovation, and Universities grant AYA2016--78994--P. We are particularly grateful to the staff of the Yebes Observatory for their extensive and invaluable help during these long observations.
\end{acknowledgements}

\newpage
\appendix

\clearpage

\section{Polarization effects} \label{appenMaserPol}

In this section, we show an example of the analysis of the circular and linear polarization evolution at 7 and 3\,mm respectively. In Fig.\,\ref{rxboo_aug2016_V1V2_RL_comp}, we compare the circular polarization (right and left) in 7\,mm observations. The time evolution of the vertical and horizontal polarizations at 3\,mm for a particular LST range is shown in Fig.\,\ref{rxboo_aug2017_V1_HV_comp}.

 \begin{figure}[htpb]
   \centering{\resizebox{9.0cm}{!}{
   \includegraphics{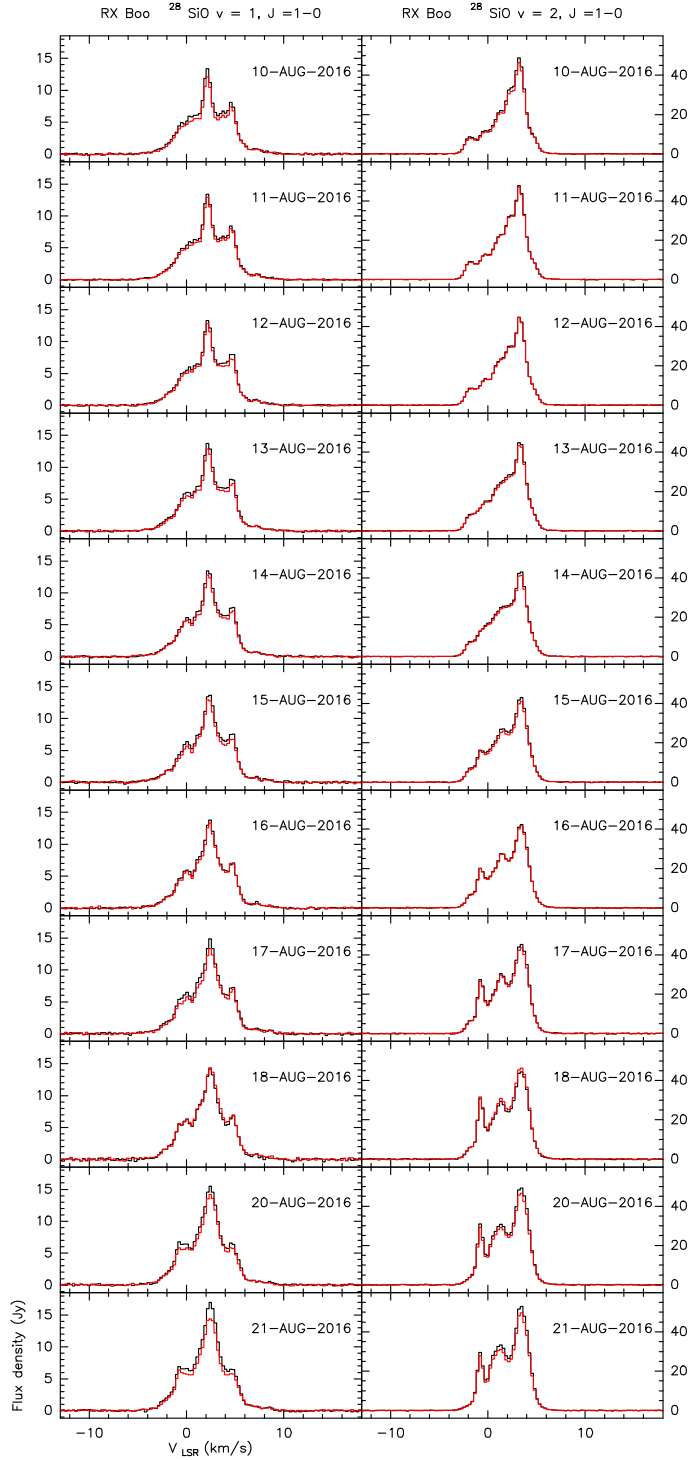}}}
   \caption{Time evolution of the right (black) and left (red) circular polarization of the $^{28}$SiO $v$=1 $J$=1--0 (\textit{right}) and $^{28}$SiO $v$=2 $J$=1--0 (\textit{left}) lines observed in RX\,Boo during our August 2016 observing run.}
    \label{rxboo_aug2016_V1V2_RL_comp}%
    \end{figure} 

 \begin{figure}[h]
   \centering{\resizebox{9.0cm}{!}{
   \includegraphics{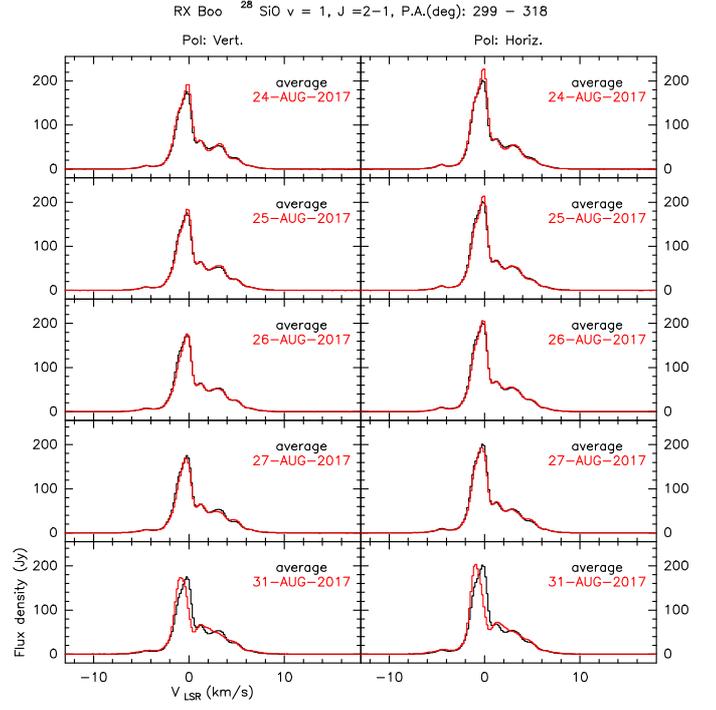}}}
   \caption{Time evolution of the vertical (left) and horizontal (rigth) linear polarization of the $^{28}$SiO $v$=1 $J$=2--1 line observed in RX\,Boo during our August 2016 observing run in a LST range between 299 and 318 degrees. We compare the average spectrum of each polarization (black) with the individual spectra (red).}
    \label{rxboo_aug2017_V1_HV_comp}%
    \end{figure}

\clearpage

\clearpage
\section{semi-regular variables: Observations} \label{appenSR}
Here, we present the observations of RX\,Boo in our August 2016, September 2016, and March 2017 observational runs (Figs.\,\ref{rxboo_aug2016_V1V2_RL_run}, \ref{rxboo_sep2016_V1V2_RL_run}, and \ref{rxboo_mar2017_V1V2_RL_run}). We also show the time evolution of the intensity of several spectral components at different velocities in Figs.\,\ref{rxboo_aug2016_v1_1-0_var}--\ref{rxboo_mar2017_v2_1-0_var}.

\begin{figure}[h]
   \centering{\resizebox{9.0cm}{!}{
   \includegraphics{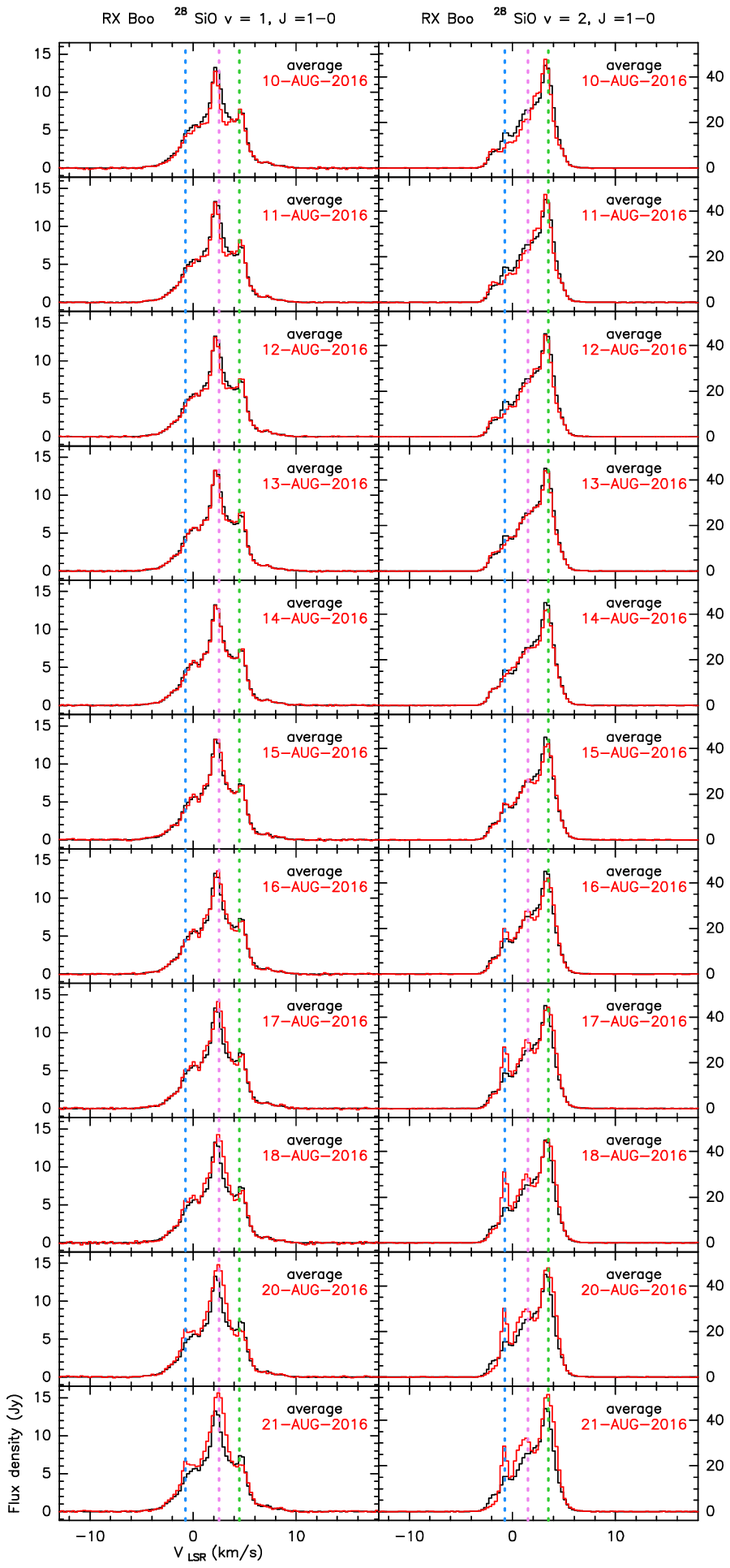}}}
   \caption{Time evolution of the $^{28}$SiO $v$=1 $J$=1--0 (\textit{rigth}) and $^{28}$SiO $v$=2 $J$=1--0 (\textit{left}) lines observed in RX\,Boo during our August 2016 observing run. The average spectra are plotted in black and the individual spectra are in red. The dashed vertical lines indicate the selected spectral components for which our statistical analysis is performed. }
    \label{rxboo_aug2016_V1V2_RL_run}%
\end{figure}

\begin{figure}[h]
   \centering{\resizebox{9.0cm}{!}{
   \includegraphics{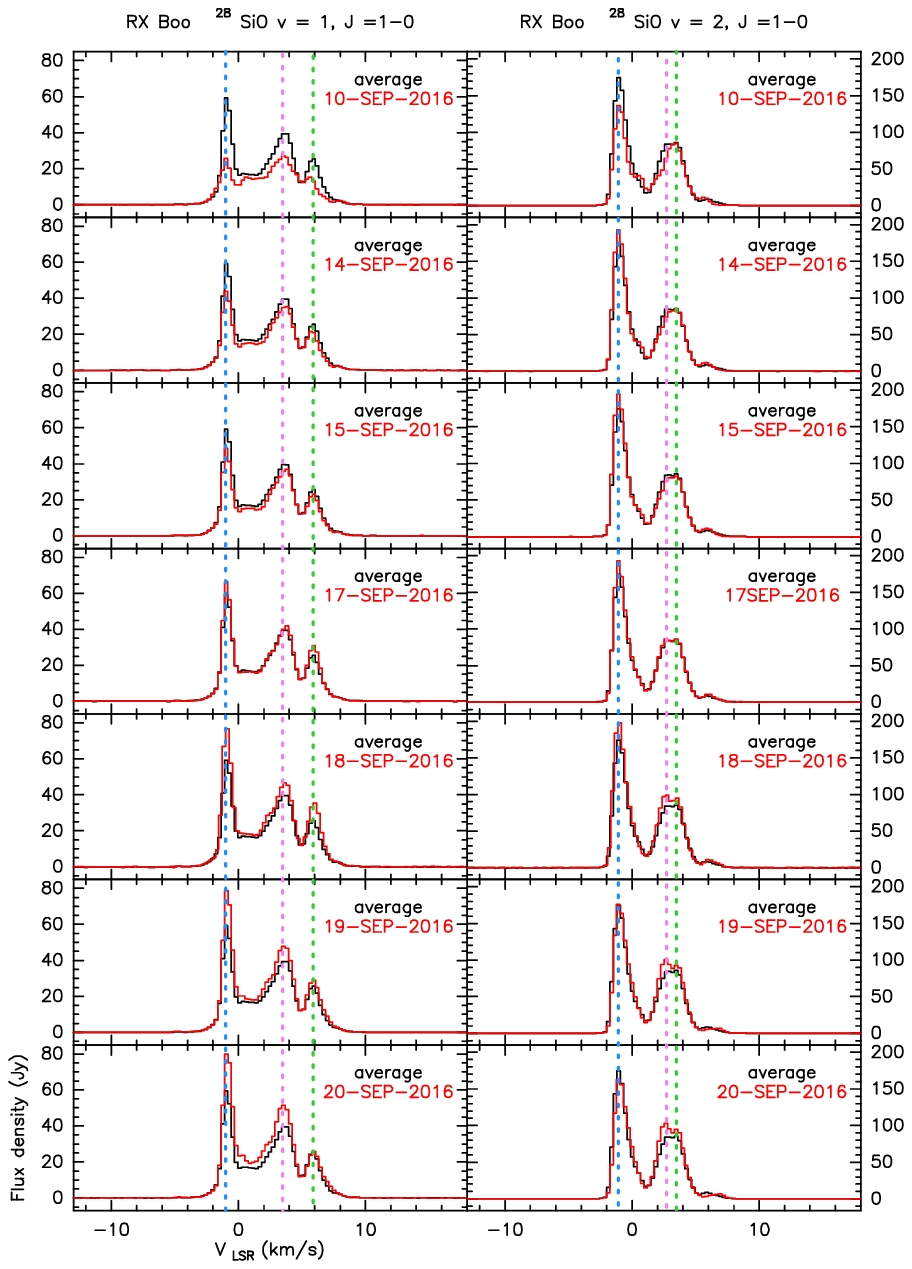}}}
   \caption{Same as Fig.\,\ref{rxboo_aug2016_V1V2_RL_run}, but for our September 2016 observing run.}
    \label{rxboo_sep2016_V1V2_RL_run}%
\end{figure}
    
\begin{figure}[h]
   \centering{\resizebox{9.0cm}{!}{
   \includegraphics{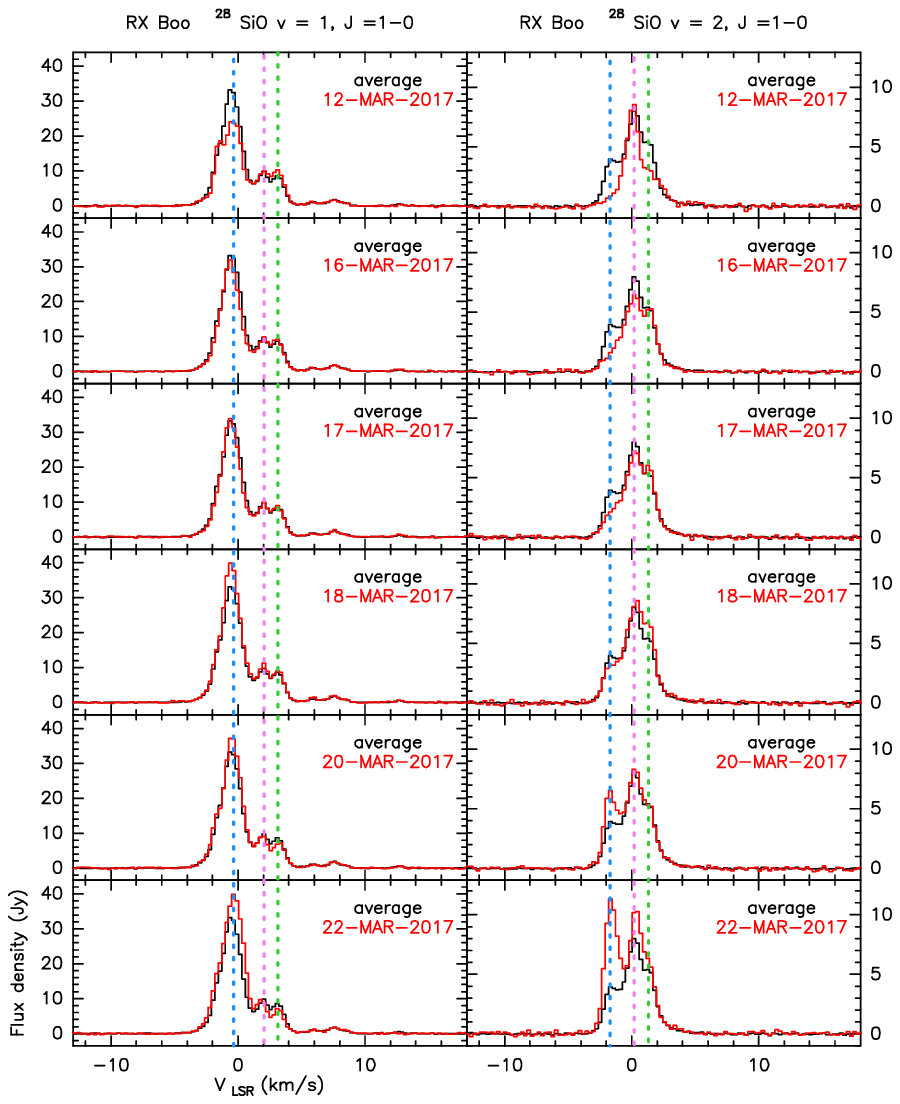}}}
   \caption{Same as Fig.\,\ref{rxboo_aug2016_V1V2_RL_run}, but for our March 2017 observing run.}
    \label{rxboo_mar2017_V1V2_RL_run}%
\end{figure}

\begin{figure}[h]
   \centering{\resizebox{8.0cm}{!}{
   \includegraphics{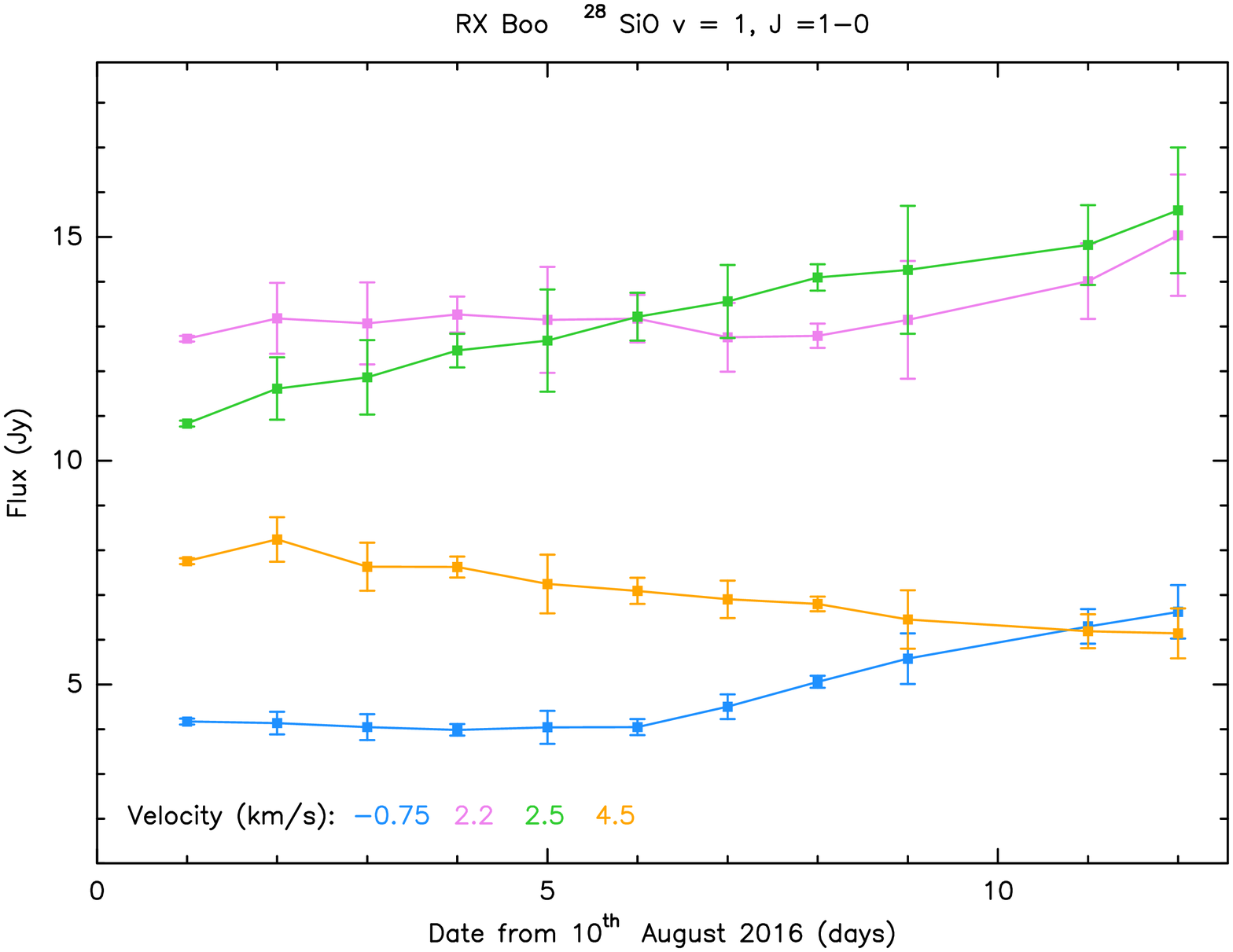}}}
   \caption{Time evolution of the $^{28}$SiO $v$=1 $J$=1--0 spectral features observed in RX\,Boo during our August 2016 observing run. The
different colors indicate the velocity of each spectral feature.}
    \label{rxboo_aug2016_v1_1-0_var}%
\end{figure}

\begin{figure}[h]
   \centering{\resizebox{8.0cm}{!}{
   \includegraphics{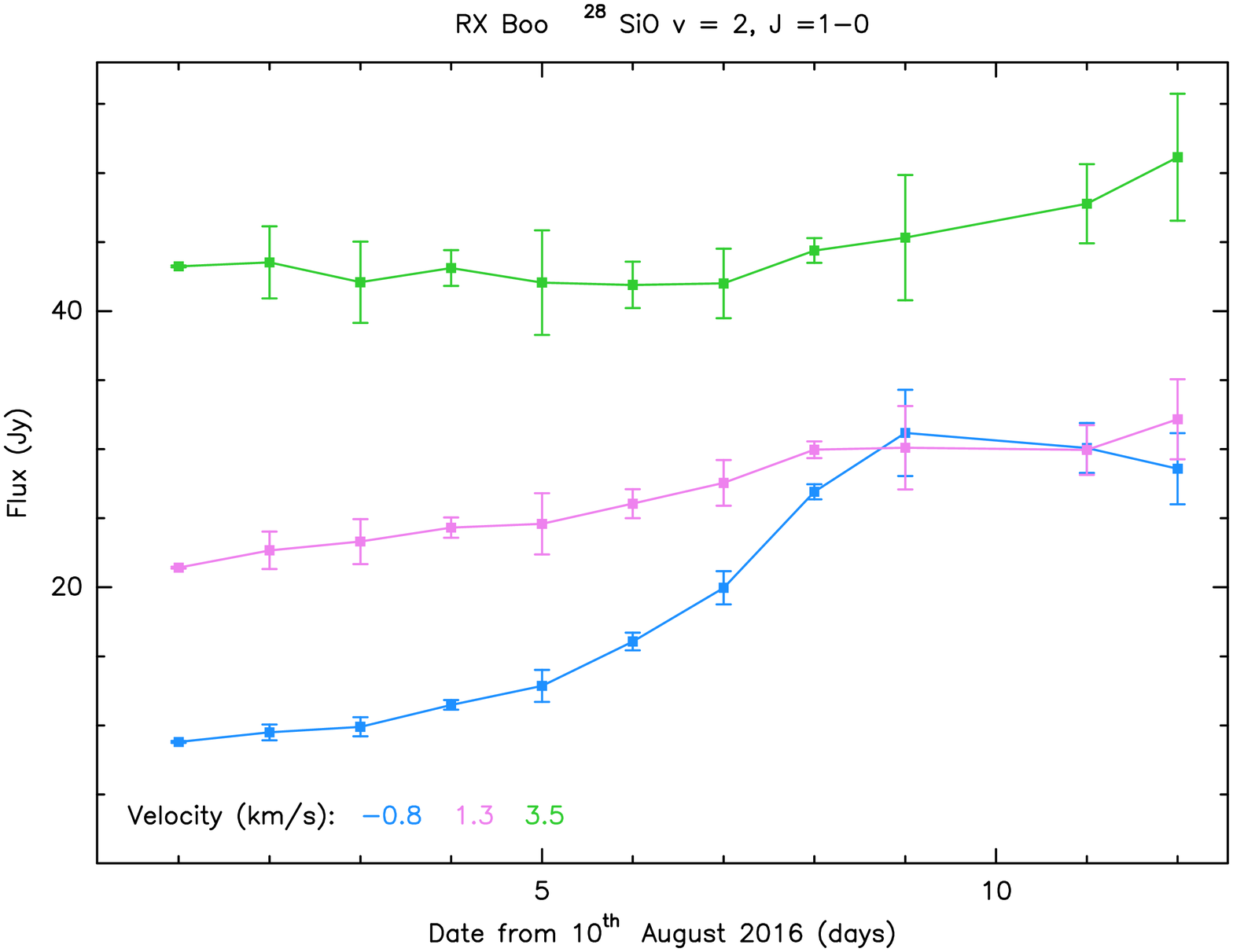}}}
   \caption{Time evolution of the $^{28}$SiO $v$=2 $J$=1--0 spectral features observed in RX\,Boo during our August 2016 observing run. The
different colors indicate the velocity of each spectral feature.}
    \label{rxboo_aug2016_v2_1-0_var}%
\end{figure}

\begin{figure}[h]
   \centering{\resizebox{8.0cm}{!}{
   \includegraphics{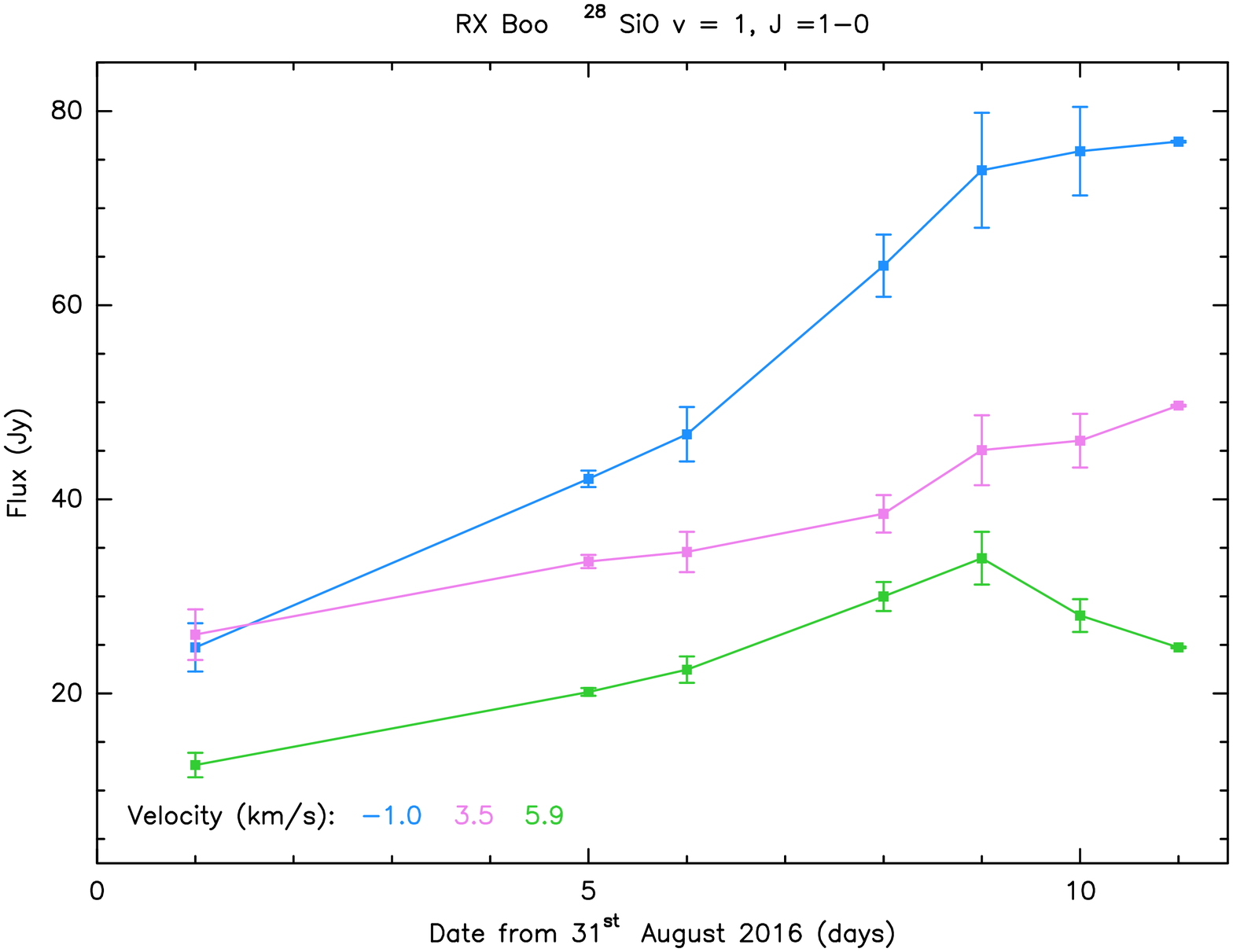}}}
   \caption{Same as Fig.\,\ref{rxboo_aug2016_v1_1-0_var}, but for our September 2016 observing run.}
    \label{rxboo_sep2016_v1_1-0_var}%
\end{figure}

\begin{figure}[h]
   \centering{\resizebox{8.0cm}{!}{
   \includegraphics{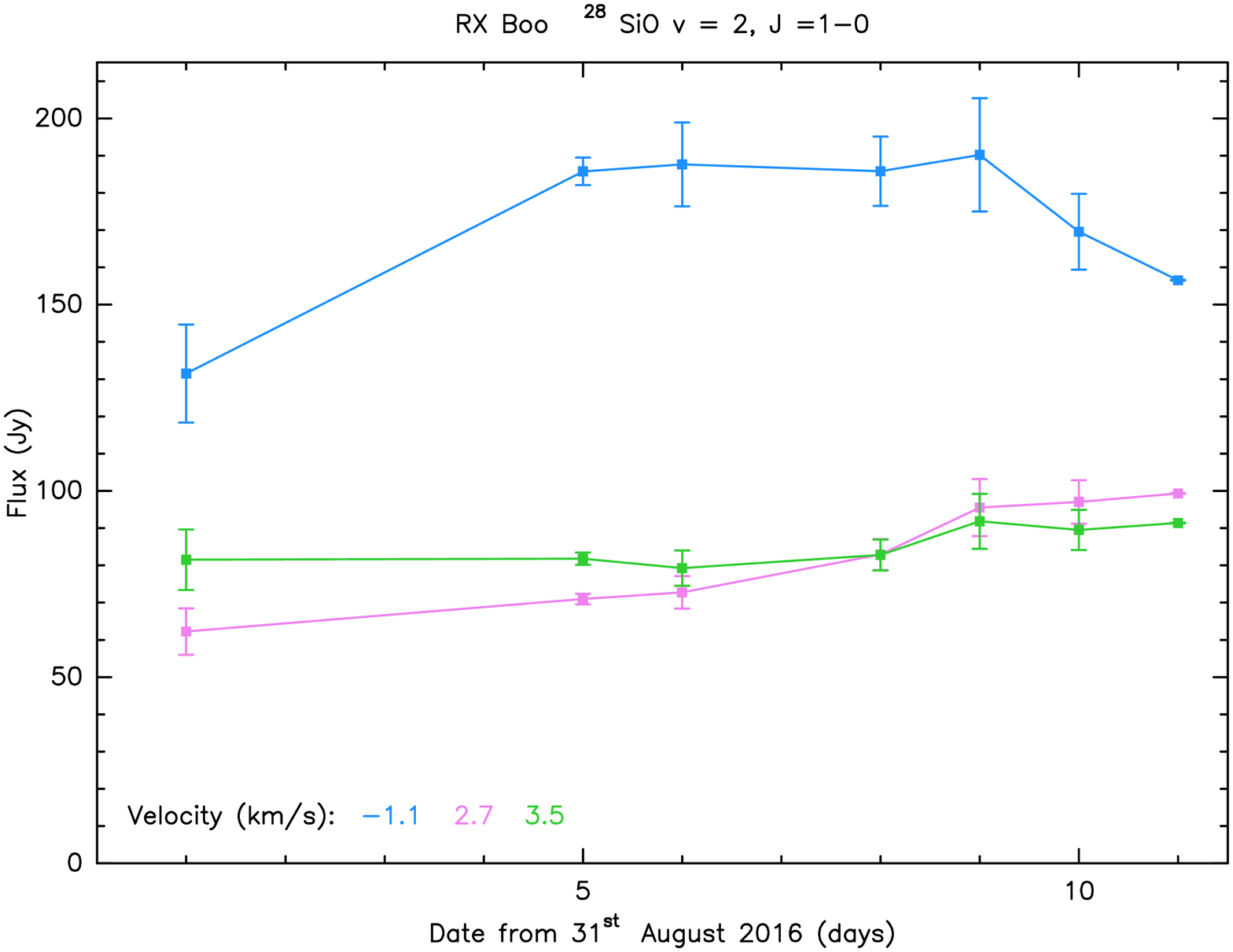}}}
   \caption{Same as Fig.\,\ref{rxboo_aug2016_v2_1-0_var}, but for our September 2016 observing run.}
    \label{rxboo_sep2016_v2_1-0_var}%
\end{figure}

\begin{figure}[h]
   \centering{\resizebox{8.0cm}{!}{
   \includegraphics{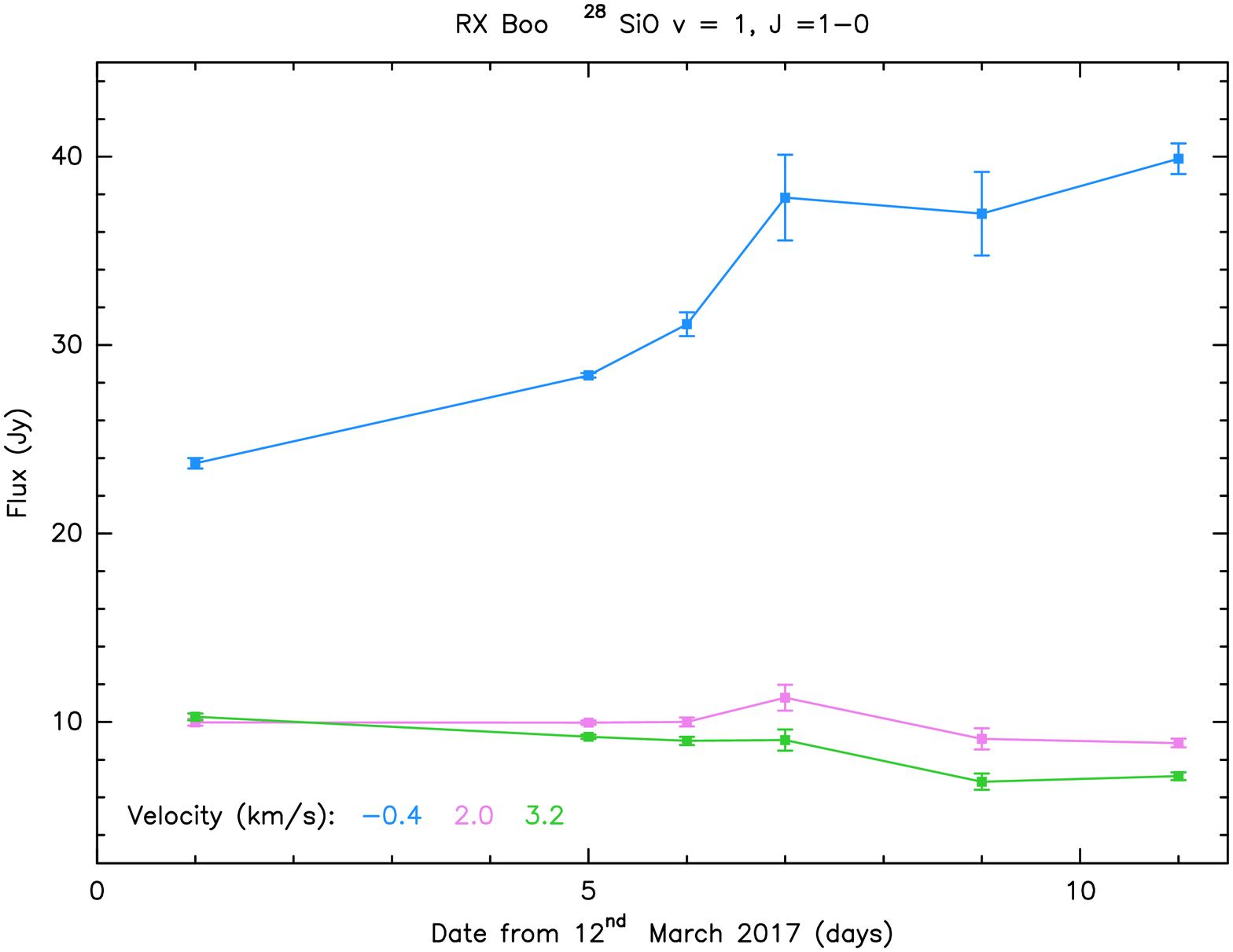}}}
   \caption{Same as Fig.\,\ref{rxboo_aug2016_v1_1-0_var}, but for our March 2017 observing run.}
    \label{rxboo_mar2017_v1_1-0_var}%
\end{figure}

\begin{figure}[h]
   \centering{\resizebox{8.0cm}{!}{
   \includegraphics{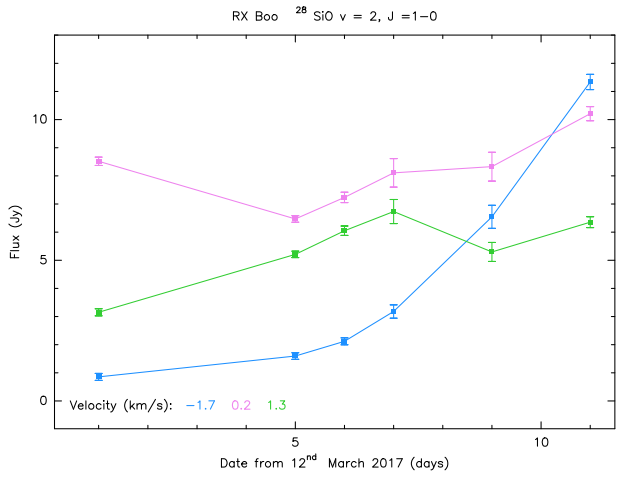}}}
   \caption{Same as Fig.\,\ref{rxboo_aug2016_v2_1-0_var}, but for our March 2017 observing run.}
    \label{rxboo_mar2017_v2_1-0_var}%
\end{figure}

\clearpage

\section{Mira-type variables: Observations} \label{appenMiras}

We present the complete set of observations of the Mira-type variables U\,Her, R\,Leo, R\,LMi, and $\chi$\,Cyg at 7 and 3\,mm.

\begin{figure}[h]
   \centering{\resizebox{8.0cm}{!}{
   \includegraphics{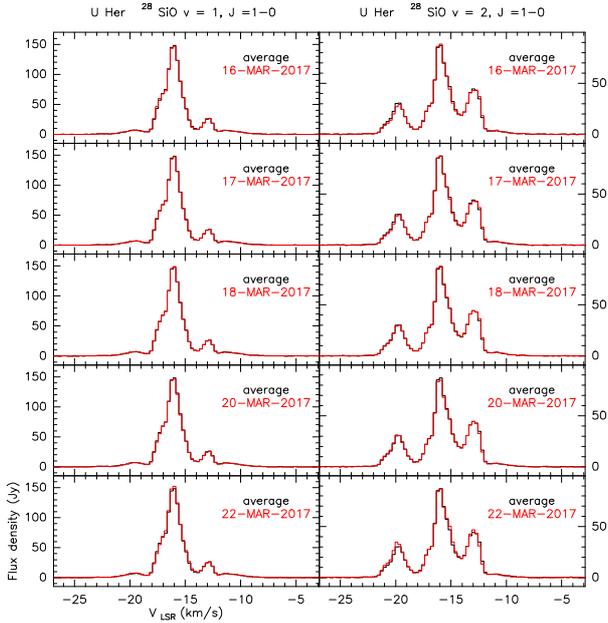}}}
   \caption{Time evolution of the $^{28}$SiO $v$=1 $J$=1--0 (\textit{right}) and $^{28}$SiO $v$=2 $J$=1--0 (\textit{left}) lines observed in U\,Her during our March 2017 observing run. The average spectra are plotted in black and the individual spectra are in red.}
    \label{uher_mar2017_v1_v2_1-0_all}%
\end{figure}

\begin{figure}[h]
   \centering{\resizebox{8.0cm}{!}{
   \includegraphics{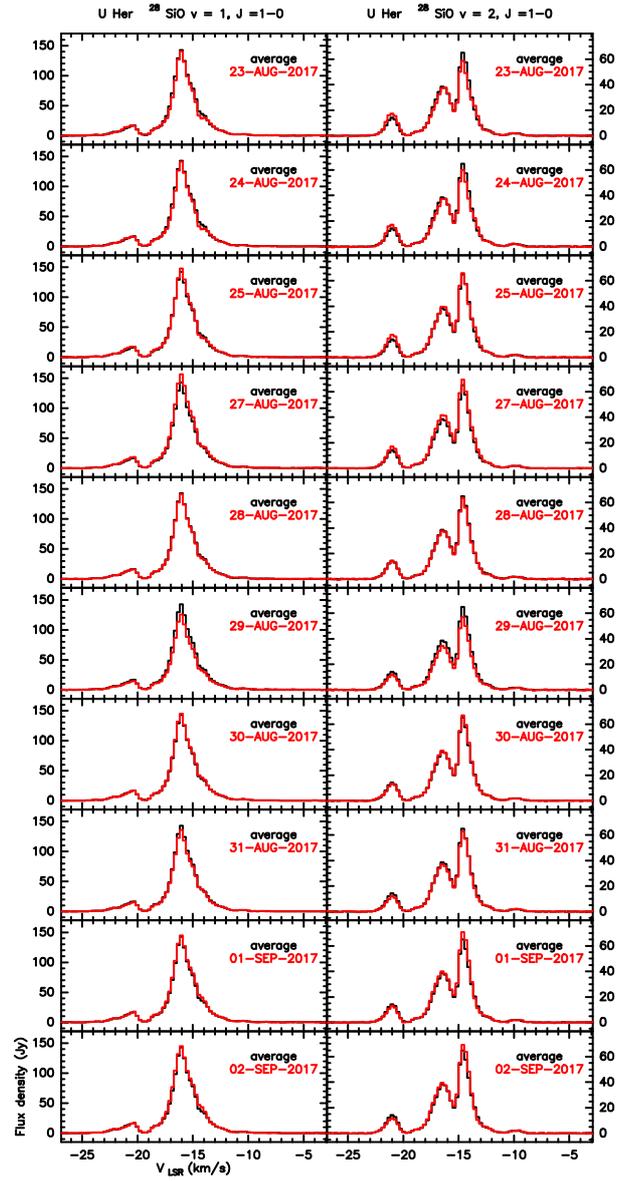}}}
   \caption{Same as Fig.\,\ref{uher_mar2017_v1_v2_1-0_all}, but for our August--September 2017 observing run. }
    \label{uher_aug2017_v1_v2_1-0_all}%
\end{figure}

\begin{figure}[h]
   \centering{\resizebox{8.0cm}{!}{
   \includegraphics{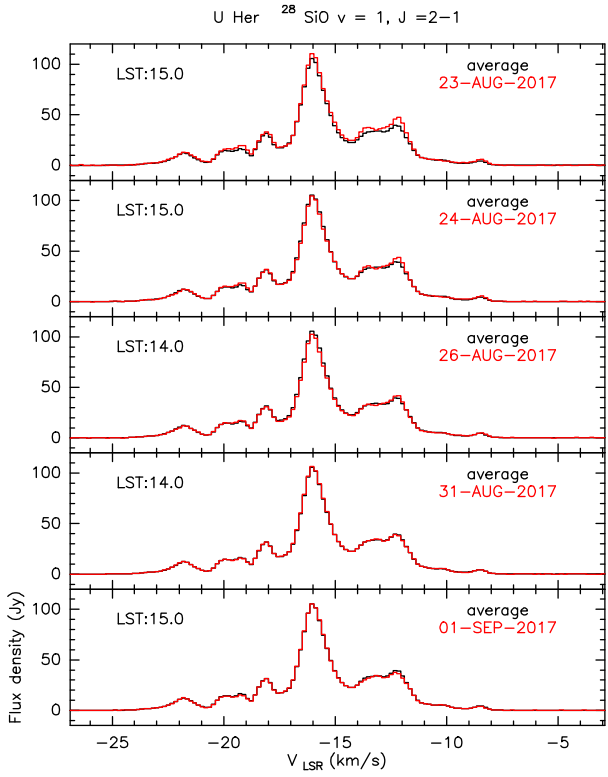}}}
   \caption{Time evolution of the $^{28}$SiO $v$=1 $J$=2--1 line observed in U\,Her during our August--September 2017 observing run. The average spectra are plotted in black and the individual spectra are in red. The dashed vertical lines indicate the selected maser components for which our statistical analysis is performed.}
    \label{uher_aug2017_v1_2-1_all}%
\end{figure}

\begin{figure}[h]
   \centering{\resizebox{8.0cm}{!}{
   \includegraphics{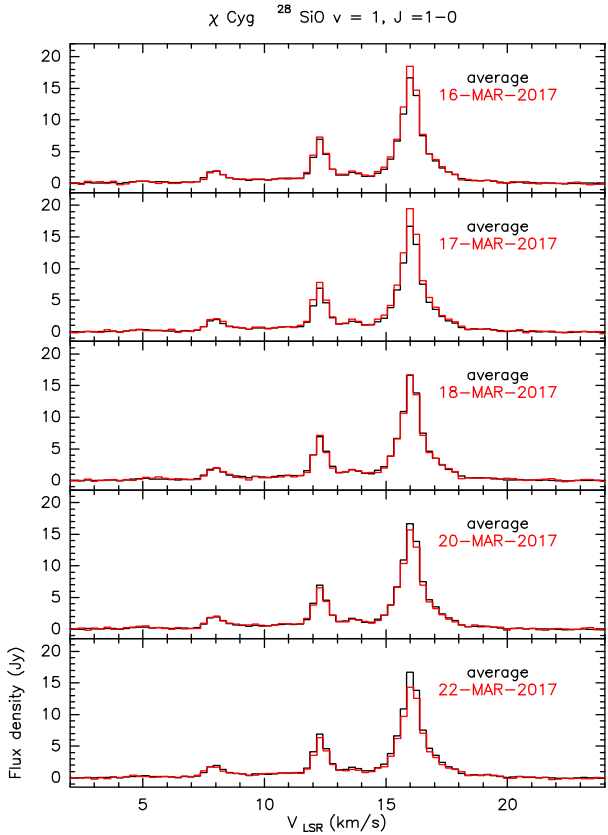}}}
   \caption{Same as Fig.\,\ref{uher_mar2017_v1_v2_1-0_all}, but for $\chi$\,Cyg.}
    \label{chicyg_mar2017_v1_1-0_all}%
\end{figure}

\begin{figure}[h]
   \centering{\resizebox{8.0cm}{!}{
   \includegraphics{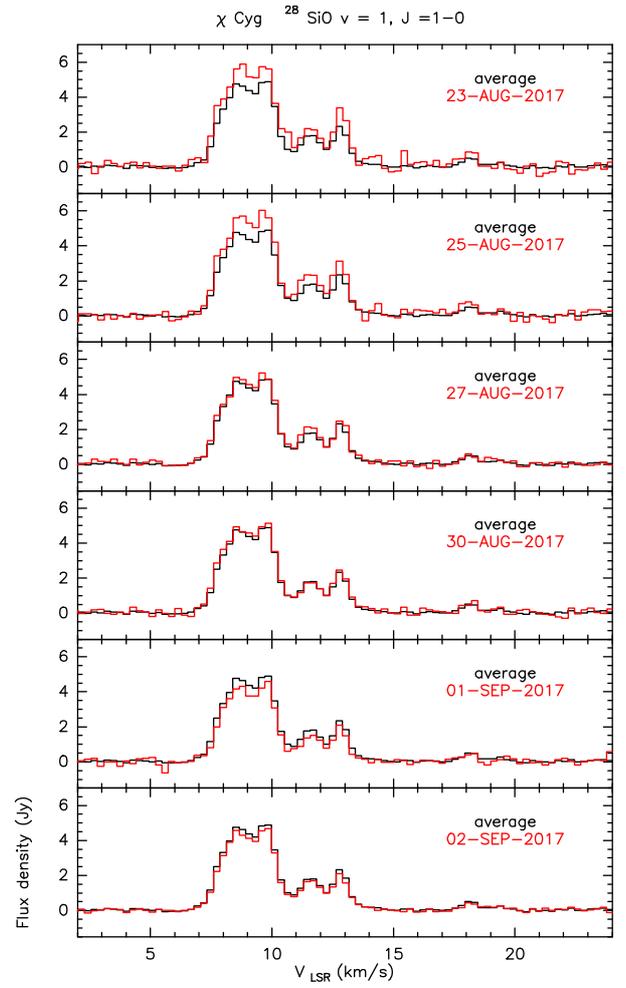}}}
   \caption{Same as Fig.\,\ref{uher_mar2017_v1_v2_1-0_all}, but for $\chi$\,Cyg in August--September 2017.}
    \label{chicyg_aug2017_v1_1-0_all}%
\end{figure}

\begin{figure}[h]
   \centering{\resizebox{8.0cm}{!}{
   \includegraphics{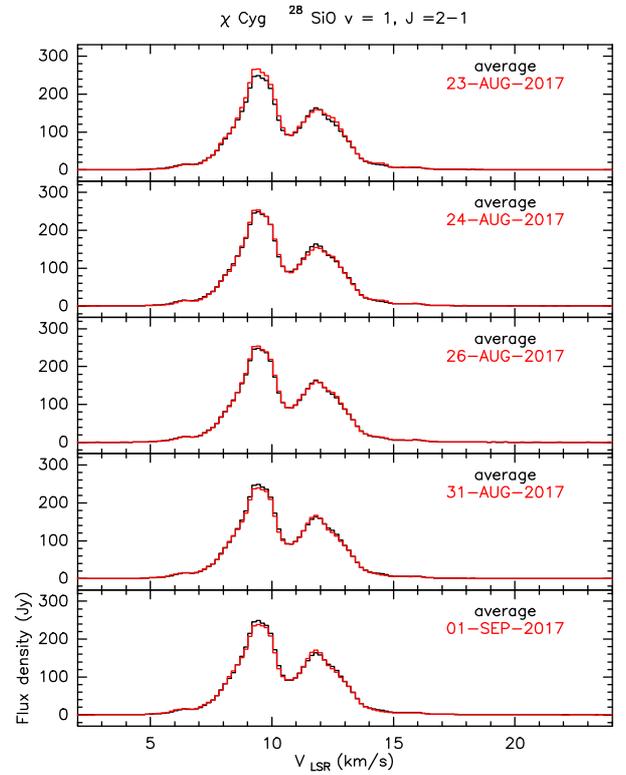}}}
   \caption{Same as Fig.\,\ref{uher_aug2017_v1_2-1_all}, but for $\chi$\,Cyg. }
    \label{chicyg_aug2017_v1_2-1_all}%
\end{figure}

\begin{figure}[h]
   \centering{\resizebox{8.0cm}{!}{
   \includegraphics{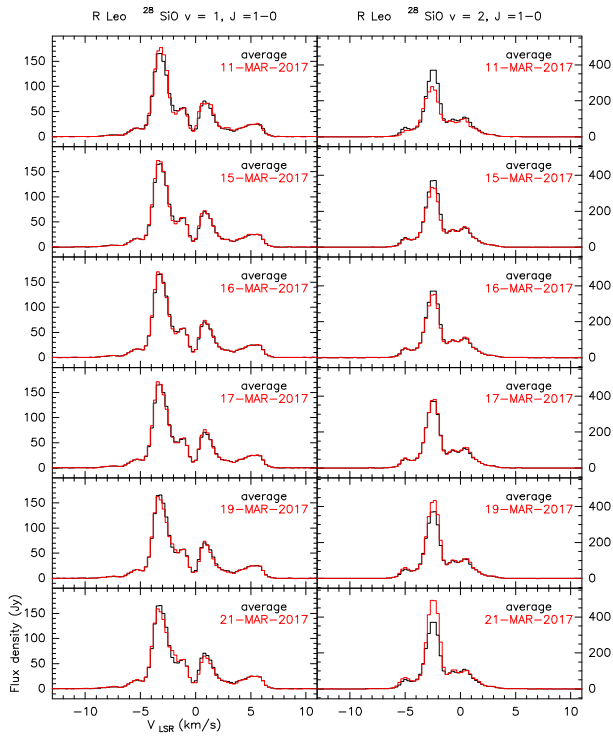}}}
   \caption{Same as Fig.\,\ref{uher_mar2017_v1_v2_1-0_all}, but for R\,Leo.}
    \label{rleo_mar2017_v1_1-0_all}%
\end{figure}

\begin{figure}[h]
   \centering{\resizebox{8.0cm}{!}{
   \includegraphics{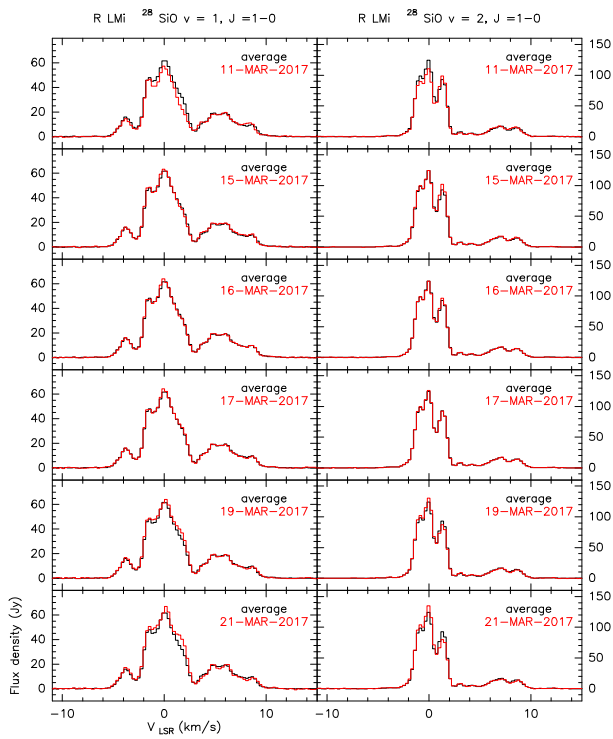}}}
   \caption{Same as Fig.\,\ref{uher_mar2017_v1_v2_1-0_all}, but for R\,LMi.}
    \label{rlmi_mar2017_v1_1-0_all}%
\end{figure}

\clearpage

\section{Further statistical analysis of the variation rate} \label{appenStat}

The time evolution of the maser features in RX\,Boo and RT\,Vir can be
studied applying well-known statistical methods such as the calculation
of the autocorrelation function (ACF) and the analysis of the frequency with
which representative variation values appear.

\subsection{Autocorrelation function}
Here we show the ACF  derived for all maser features detected in the $v$=1 $J$=1-0 line during the August 2017 run as an example of this procedure. The result of the analysis can be seen in Fig.\,\ref{acf_rxboo_aug2017_v1_1-0}. The red lines indicate significance levels for 95$\%$ confidence.

We reiterate that for a time series like $X_1$, $X_2$...$X_N$, the covariance is defined as \citep[see][]{venables02}:
\begin{equation}
c_{\tau}=\frac{1}{N}\sum_{\lambda=max(1,-\tau)}^{min(N-\tau,N)} \left[X_{s+t}-X_{mean}\right]\left[X_{s}-X_{mean}\right]
\label{eq.cov}
.\end{equation}

The autocorrelation function is then

\begin{equation}
ACF(\tau)=\frac{c_\tau}{c_0}
\label{eq.acf}
.\end{equation}

Furthermore, the significance level is given by $1.96\,/\sqrt{N}$.

The only phase lag in the August 2017 run  that shows a significant autocorrelation, besides zero, is one, which corresponds to the minimum sampling between observations. Its significance simply  means that variations do not occur on much shorter time periods. The remaining phase lags do not show significant ACF values. The negative values found for a lag of about 7 days may indicate an anti-correlation, but they are really not significant. For the other observational runs, the ACF analysis is still less concluding.

We conclude that the total observing period of about 10 days is not long enough to detect eventual periodical variations in our case. The ACF is not found to be very useful in the statistical analysis of the observed flux density variations.

\begin{figure}[h]
   \centering{\resizebox{9.0cm}{!}{
   \includegraphics{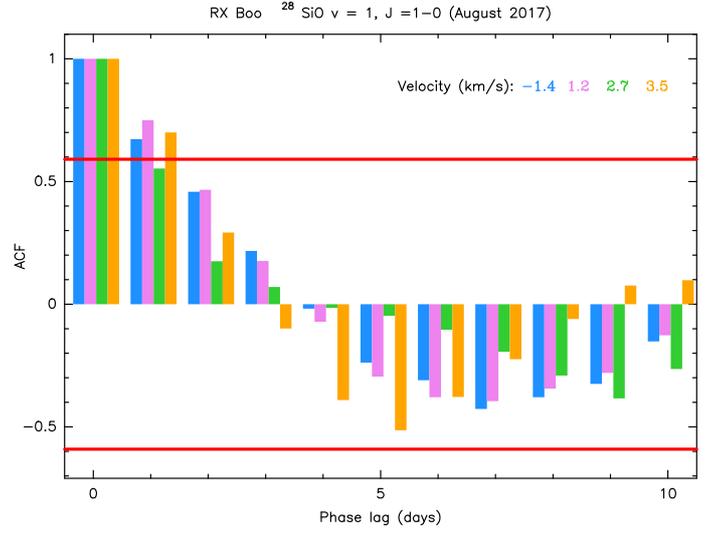}}}
   \caption{Autocorrelation function of the spectral features at -1.4, 1.2, 2.7, and 3.5 \,kms$^{-1}$ for the $^{28}$SiO $v$=1 $J$=1--0 line observed in RX\,Boo for our August 2017 observational run. The red horizontal lines indicate the significance level for 95$\%$  confidence.}
    \label{acf_rxboo_aug2017_v1_1-0}%
\end{figure}

\subsection{Frequency of variation rate values}

In order to quantify the variation rate of the different maser features, we calculate the number of times that the variation of a single spectral feature is higher than a certain percentage for the total set of experiments we performed. We define this quantity as $\rho \left(\tau,\Delta\right)$, depending on the time spacing ($\tau$) and the variation rate ($\Delta$):
\begin{equation}
\rho\left(\tau,\Delta\right)=\frac{\sum_{i,j} X\left(j-i=\tau,\Delta\right)}{N}
\label{eq.1}
,\end{equation}
where $N$ is the total number of experiments, and $X\left(j-i,\Delta\right)$ is defined as
\begin{equation}
X\left(j-i,\Delta\right) = \left\lbrace
\begin{array}{ll}
1 & \textup{if } \delta([j-i])>\Delta\\
0 & \textup{if } \delta([j-i])< \Delta
\end{array}
\right.
\label{eq.2}
.\end{equation}

Here $\delta([j-i])$ is the maximum flux density variation in a given time interval:

\begin{equation}
\delta([j,i])= \frac{F_{max}[j,i]-F_{min}[j,i]}{F_{mean}[j,i]}\
\label{eq.3}
,\end{equation}

where $F_{max}[j,i]$, $F_{min}[j,i]$ and $F_{mean}[j,i]$ are the flux density maximum, minimum, and mean in the interval between the days $i$ and $j$.

In the following tables, we show the results derived from the analysis of the variation rate on the semi-regular variables RX\,Boo and RT\,Vir. \\

\begin{table}[htbp]
\caption{Statistical results for the $^{28}$SiO $v$=1 $J$=1--0 line observed in RX\,Boo for our August 2016 observational run. The analysis is performed for the variation rates of 10\% and 40\% at various velocities.}
\centering
\begin{spacing}{1.5}
\begin{tabular}{ c | c c c c  }
\noalign{\smallskip}
\multicolumn{5}{c}{August 2016 ~~ $v$=1 $J$=1--0} \\ \hline \hline
\multirow{2}{*}{Spectral} & \multicolumn{4}{c}{Gap (days)} \\
 component (kms$^{-1}$)  & 1 & 2 & 4 & 7 \\ \hline \hline
 & \multicolumn{4}{c}{Variation rate: 10\%} \\ \hline 
        -0.75 & 22\% & 38\% & 71\% & 100\% \\ \hline
         2.20 & 0\% & 0\% & 14\% & 25\% \\ \hline
         2.50 & 0\% & 0\% & 86\% & 100\% \\ \hline
         4.50 & 0\% & 0\% & 86\% & 100\% \\ \hline
  \hline
 & \multicolumn{4}{c}{Variation rate: 40\%} \\ \hline
        -0.75 & 0\% & 0\% & 0\% & 50\% \\ \hline
         2.20 & 0\% & 0\% & 0\% & 0\% \\ \hline
         2.50 & 0\% & 0\% & 0\% & 0\% \\ \hline
         4.50 & 0\% & 0\% & 0\% & 0\% \\ \hline
 
\end{tabular} 
\end{spacing}
\label{tab.var10-40V1aug16_1-0}
\end{table}

\begin{table}[htbp]
\caption{Same as Table\,\ref{tab.var10-40V1aug16_1-0}, but for the $^{28}$SiO $v$=2 $J$=1--0 line.}
\centering
\begin{spacing}{1.5}
\begin{tabular}{ c | c c c c  }
\noalign{\smallskip}
\multicolumn{5}{c}{August 2016 ~~ $v$=2 $J$=1--0} \\ \hline \hline
\multirow{2}{*}{Spectral} & \multicolumn{4}{c}{Gap (days)} \\
component (kms$^{-1}$)  & 1 & 2 & 4 & 7 \\ \hline \hline
 & \multicolumn{4}{c}{Variation rate: 10\%} \\ \hline 
        -0.80 & 44\% & 75\% & 100\% & 100\% \\ \hline
         1.30 & 0\% & 25\% & 71\% & 100\% \\ \hline
         3.50 & 0\% & 0\% & 29\% & 50\% \\ \hline
  \hline
 & \multicolumn{4}{c}{Variation rate: 40\%} \\ \hline
        -0.80 & 0\% & 25\% & 57\% & 100\% \\ \hline
         1.30 & 0\% & 0\% & 0\% & 0\% \\ \hline
         3.50 & 0\% & 0\% & 0\% & 0\% \\ \hline
\end{tabular} 
\end{spacing}
\label{tab.var10-40V2aug16_1-0}
\end{table}

\begin{table}[htbp]
\caption{Same as Table\,\ref{tab.var10-40V1aug16_1-0}, but for September 2016 observational run.}
\centering
\begin{spacing}{1.5}
\begin{tabular}{ c | c c c c  }
\noalign{\smallskip}
\multicolumn{5}{c}{September 2016 ~~ $v$=1 $J$=1--0} \\ \hline \hline
\multirow{2}{*}{Spectral} & \multicolumn{4}{c}{Gap (days)} \\
component (kms$^{-1}$)  & 1 & 2 & 4 & 7 \\ \hline \hline
 & \multicolumn{4}{c}{Variation rate: 10\%} \\ \hline 
        -1.00 & 25\% & 67\% & 100\% & 100\% \\ \hline
         3.50 & 25\% & 100\% & 100\% & 100\% \\ \hline
         5.90 & 75\% & 100\% & 100\% & 100\% \\ \hline
  \hline
 & \multicolumn{4}{c}{Variation rate: 40\%} \\ \hline
        -1.00 & 0\% & 0\% & 67\% & 100\% \\ \hline
         3.50 & 0\% & 0\% & 0\% & 100\% \\ \hline
         5.90 & 0\% & 0\% & 67\% & 100\% \\ \hline
\end{tabular} 
\end{spacing}
\label{tab.var10-40V1sep16_1-0}
\end{table}

\begin{table}[htbp]
\caption{Same as Table\,\ref{tab.var10-40V2aug16_1-0}, but for September 2016 observational run.}
\centering
\begin{spacing}{1.5}
\begin{tabular}{ c | c c c c  }
\noalign{\smallskip}
\multicolumn{5}{c}{September 2016 ~~ $v$=2 $J$=1--0} \\ \hline \hline
\multirow{2}{*}{Spectral} & \multicolumn{4}{c}{Gap (days)} \\
component (kms$^{-1}$)  & 1 & 2 & 4 & 7 \\ \hline \hline
 & \multicolumn{4}{c}{Variation rate: 10\%} \\ \hline 
        -1.10 & 25\% & 67\% & 67\% & 100\% \\ \hline
         2.70 & 25\% & 67\% & 100\% & 100\% \\ \hline
         3.50 & 25\% & 0\% & 0\% & 0\% \\ \hline
  \hline
 & \multicolumn{4}{c}{Variation rate: 40\%} \\ \hline
        -1.00 & 0\% & 0\% & 0\% & 0\% \\ \hline
         2.70 & 0\% & 0\% & 0\% & 0\% \\ \hline
         3.50 & 0\% & 0\% & 0\% & 0\% \\ \hline
\end{tabular} 
\end{spacing}
\label{tab.var10-40V2sep16_1-0}
\end{table}

\begin{table}[htbp]
\caption{Same as Table\,\ref{tab.var10-40V1aug16_1-0}, but for March 2017 observational run.}
\centering
\begin{spacing}{1.5}
\begin{tabular}{ c | c c c c  }
\noalign{\smallskip}
\multicolumn{5}{c}{March 2017 ~~ $v$=1 $J$=1--0} \\ \hline \hline
\multirow{2}{*}{Spectral} & \multicolumn{4}{c}{Gap (days)} \\
component (kms$^{-1}$)  & 1 & 2 & 4 & 7 \\ \hline \hline
 & \multicolumn{4}{c}{Variation rate: 10\%} \\ \hline 
   -0.40 & 50\% & 33\% & 83\% & 100\% \\ \hline
    2.00 & 50\% & 66\% & 67\% & 100\% \\ \hline
    3.20 & 0\% & 33\% & 100\% & 100\% \\ \hline
  \hline
 & \multicolumn{4}{c}{Variation rate: 40\%} \\ \hline
   -0.40 & 0\% & 0\% & 0\% & 75\% \\ \hline
    2.00 & 0\% & 0\% & 0\% & 0\% \\ \hline
    3.20  & 0\% & 0\% & 0\% & 0\% \\ \hline
\end{tabular} 
\end{spacing}
\label{tab.var10-40V1mar17_1-0}
\end{table}

\begin{table}[htbp]
\caption{Same as Table\,\ref{tab.var10-40V2aug16_1-0}, but for March 2017 observational run.}
\centering
\begin{spacing}{1.5}
\begin{tabular}{ c | c c c c  }
\noalign{\smallskip}
\multicolumn{5}{c}{March 2017 ~~ $v$=2 $J$=1--0} \\ \hline \hline
\multirow{2}{*}{Spectral} & \multicolumn{4}{c}{Gap (days)} \\
component (kms$^{-1}$)  & 1 & 2 & 4 & 7 \\ \hline \hline
 & \multicolumn{4}{c}{Variation rate: 10\%} \\ \hline 
   -1.70 & 100\% & 100\% & 100\% & 100\% \\ \hline
    0.20 & 50\% & 67\% & 100\% & 100\% \\ \hline
    1.30 & 100\% & 100\% & 100\% & 100\% \\ \hline
  \hline
 & \multicolumn{4}{c}{Variation rate: 40\%} \\ \hline
   -1.70 & 0\% & 67\% & 67\% & 100\% \\ \hline
    0.20 & 0\% & 0\% & 0\% & 63\% \\ \hline
    1.30  & 0\% & 0\% & 0\% & 75\% \\ \hline
\end{tabular} 
\end{spacing}
\label{tab.var10-40V2mar17_1-0}
\end{table}
    
 \begin{table}[htbp]
\caption{Same as Table\,\ref{tab.var10-40V1aug16_1-0}, but for August--September 2017 observational run.}
\centering
\begin{spacing}{1.5}
\begin{tabular}{ c | c c c c  }
\noalign{\smallskip}
\multicolumn{5}{c}{August 2017 ~~ $v$=1 $J$=1--0} \\ \hline \hline
\multirow{2}{*}{Spectral} & \multicolumn{4}{c}{Gap (days)} \\
component (kms$^{-1}$)  & 1 & 2 & 4 & 7 \\ \hline \hline
 & \multicolumn{4}{c}{Variation rate: 10\%} \\ \hline 
   -1.40 & 20\% & 56\% & 100\% & 100\% \\ \hline
    1.20 & 20\% & 56\% & 71\% & 100\% \\ \hline
    2.70 & 20\% & 44\% & 100\% & 100\% \\ \hline
    3.50 & 30\% & 56\% & 100\% & 100\% \\ \hline
  \hline
 & \multicolumn{4}{c}{Variation rate: 40\%} \\ \hline
   -1.40 & 0\% & 0\% & 0\% & 0\% \\ \hline
    1.20 & 0\% & 0\% & 14\% & 75\% \\ \hline
    2.70 & 0\% & 11\% & 29\% & 50\% \\ \hline
    3.50 & 0\% & 11\% & 29\% & 100\% \\ \hline
\end{tabular} 
\end{spacing}
\label{tab.var10-40V1aug17_1-0}
\end{table}

\begin{table}[htbp]
\caption{Same as Table\,\ref{tab.var10-40V2aug16_1-0}, but for August--September 2017 observational run.}
\centering
\begin{spacing}{1.5}
\begin{tabular}{ c | c c c c  }
\noalign{\smallskip}
\multicolumn{5}{c}{August 2017 ~~ $v$=2 $J$=1--0.} \\ \hline \hline
\multirow{2}{*}{Spectral} & \multicolumn{4}{c}{Gap (days)} \\
component (kms$^{-1}$)  & 1 & 2 & 4 & 7 \\ \hline \hline
 & \multicolumn{4}{c}{Variation rate: 10\%} \\ \hline 
    2.50 & 80\% & 100\% & 100\% & 100\% \\ \hline
    3.20 & 70\% & 100\% & 100\% & 100\% \\ \hline
    6.10 & 50\% & 67\% & 100\% & 100\% \\ \hline
    8.30 & 70\% & 89\% & 100\% & 100\% \\ \hline
  \hline
 & \multicolumn{4}{c}{Variation rate: 40\%} \\ \hline
    2.50 & 10\% & 22\% & 57\% & 100\% \\ \hline
    3.20 & 0\% & 0\% & 43\% & 100\% \\ \hline
    6.10 & 10\% & 22\% & 57\% & 100\% \\ \hline
    8.30 & 10\% & 44\% & 100\% & 100\% \\ \hline
\end{tabular} 
\end{spacing}
\label{tab.var10-40V2aug17_1-0}
\end{table}

\begin{table}[htbp]
\caption{Statistical results for the $^{28}$SiO $v$=1 $J$=2--1 line observed in RX\,Boo for our August--September 2017 observational run. The analysis is performed for the variation rates of 10\% and 40\% at various velocities.}
\centering
\begin{spacing}{1.5}
\begin{tabular}{ c | c c c c  }
\noalign{\smallskip}
\multicolumn{5}{c}{August 2017 ~~ $v$=1 $J$=2--1} \\ \hline \hline
\multirow{2}{*}{Spectral} & \multicolumn{4}{c}{Gap (days)} \\
component (kms$^{-1}$)  & 1 & 2 & 4 & 7 \\ \hline \hline
 & \multicolumn{4}{c}{Variation rate: 10\%} \\ \hline 
   -0.90 & 0\% & 25\% & 100\% & 100\% \\ \hline
    0.00 & 25\% & 100\% & 100\% & 100\% \\ \hline
    1.10 & 0\% & 0\% & 0\% & 0\% \\ \hline
    3.20 & 0\% & 25\% & 50\% & 100\% \\ \hline
  \hline
 & \multicolumn{4}{c}{Variation rate: 40\%} \\ \hline
   -0.90 & 0\% & 0\% & 0\% & 0\% \\ \hline
    0.00 & 0\% & 0\% & 100\% & 100\% \\ \hline
    1.10  & 0\% & 0\% & 0\% & 0\% \\ \hline
    3.20 & 0\% & 0\% & 0\% & 0\% \\ \hline
\end{tabular} 
\end{spacing}
\label{tab.var10-40V1aug17_2-1}
\end{table}


\begin{table}[htpb]
\caption{Statistical results for the $^{28}$SiO $v$=1 $J$=1--0 line observed in RT\,Vir for our November 2017 observational run. The analysis is performed for the variation rates of 10\% and 40\% at various velocities.}
\centering
\begin{spacing}{1.5}
\begin{tabular}{ c | c c c c  }
\noalign{\smallskip}
\multicolumn{5}{c}{November 2017 ~~ $v$=1 $J$=1--0.} \\ \hline \hline
\multirow{2}{*}{Spectral} & \multicolumn{4}{c}{Gap (days)} \\
component (kms$^{-1}$)  & 1 & 2 & 4 & 7 \\ \hline \hline
 & \multicolumn{4}{c}{Variation rate: 10\%} \\ \hline 
    13.2 & 100\% & 100\% & 100\% & 100\% \\ \hline
    15.6 & 50\% & 100\% & 100\% & 100\% \\ \hline
  \hline
 & \multicolumn{4}{c}{Variation rate: 40\%} \\ \hline
    13.2 & 0\% & 0\% & 0\% & 100\% \\ \hline
    15.6 & 0\% & 0\% & 0\% & 100\% \\ \hline
\end{tabular} 
\end{spacing}
\label{tab.rtvir_var10-40V1nov17_1-0}
\end{table}

\begin{table}[htpb]
\caption{Statistical results for the $^{28}$SiO $v$=2 $J$=1--0 line observed in RT\,Vir for our November 2017 observational run. The analysis is performed for the variation rates of 10\% and 40\% at various velocities.}
\centering
\begin{spacing}{1.5}
\begin{tabular}{ c | c c c c  }
\noalign{\smallskip}
\multicolumn{5}{c}{November 2017 ~~ $v$=2 $J$=1--0} \\ \hline \hline
\multirow{2}{*}{Spectral} & \multicolumn{4}{c}{Gap (days)} \\
component (kms$^{-1}$)  & 1 & 2 & 4 & 7 \\ \hline \hline
 & \multicolumn{4}{c}{Variation rate: 10\%} \\ \hline 
    16.0 & 50\% & 100\% & 100\% & 100\% \\ \hline
    17.5 & 50\% & 100\% & 100\% & 100\% \\ \hline
  \hline
 & \multicolumn{4}{c}{Variation rate: 40\%} \\ \hline
    16.0 & 0\% & 0\% & 100\% & 100\% \\ \hline
    17.50 & 0\% & 33\% & 100\% & 100\% \\ \hline
\end{tabular} 
\end{spacing}
\label{tab.rtvir_var10-40V2nov17_1-0}
\end{table}

\clearpage


\begin{thebibliography}{}
\bibitem[Alcolea et al.\,(1999)]
        {alcolea99} Alcolea, J., Pardo, J.R., Bujarrabal, V. et al. 1999, A\&AS, 139, 461

\bibitem[Balister et al.\,(1977)]
        {balister77} Balister, M., Batchelor, R. A., Haynes, R. F. et al. 1977, MNRAS, 180, 415

\bibitem[Boboltz \& Claussen\,(2004)]
        {bobcla04} Boboltz, D. A., \& Claussen, M. J.\ 2004, \apj, 608, 480 
        
\bibitem[Bujarrabal \& Alcolea\,(1991)]
        {bujarrabal91} Bujarrabal, V. \& Alcolea, J. 1991, A\&A, 251, 536
        
\bibitem[de Vicente et al.\,(2016)]
        {devicente16} de Vicente, P., Bujarrabal, V., Díaz-Pulido, A., et al. 2016, A\&A, 589, A74
        
\bibitem[Diamond et al.\,(1994)]
        {diamond94} Diamond, P. J., Kemball, A. J., Junor, W., et al. 1994, ApJ, 430, L61.
        
\bibitem[Gaia Collaboration\,(2018)]
        {gaiaColl18} Gaia Collaboration (Brown, A. G. A., et al.) 2018, A\&A, 616, A1.
        
\bibitem[González et al.\,(2003)]
        {gonzalez03} González Delgado, D., Olofsson, H., Kerschbaum, F., et al. 2003, A\&A, 411, 123

\bibitem[Habing\,(1996)]
     {habing96} Habing, H., 1996, A\&AR, Vol. 7, Issue 2, 97 %

\bibitem[Herpin et al.\,(2006)]
        {herpin06} Herpin, F., Baudry, A. Thum, C., et al. 2006, A\&A, 450, 667H

\bibitem[Hinkle\,(1978)]
        {hinkle78} Hinkle, K. H. 1978, ApJ, 220, 210H
        
\bibitem[Höfner at al.\,(2003)]
        {hofner03} Höfner, S., Gautschy-Loidl, R., Aringer, B., Jørgensen, U. G. 2003, A\&A, 399, 589
        
\bibitem[Humphreys et al.\,(2002)]
        {humphreys02} Humphreys, E. M. L., Gray, M. D., Yates, J. A., et al. 2002, A\&A, 386, 256
        
\bibitem[Kemball \& Diamond\,(1997)]
        {kemball97} Kemball, J. \& Diamond, P.J. 1997, ApJ, 481, L111
\bibitem[Kim et al.\,(2018)]
        {kim08} Kim, D.-J., Cho, S.-H., Yun, Y. et al. 2018, ApJ, 866, L19
        
\bibitem[Locket \& Elitzur\,(1992)]
    {locket92}  Lockett, P. \& Elitzur, M., 1992, ApK, 399, 704
        
\bibitem[Lucas et al.\,(1992)]
        {lucas92} Lucas, R., Bujarrabal, V., Guilloteau, S., et al. 1992, A\&A, 262, 491
        
\bibitem[McIntosh \& Indermuehle\,(2015)]
        {mcintosh15} McIntosh, G. \& Indermuehle, B. 2015, AJ, 149, 100M

\bibitem[Pardo et al.\,(1998)]
    {pardo98} Pardo, J. R., Cernicharo, J., González-Alfonso, E., et al. 1998, A\&A, 329, 219

 \bibitem[Pardo et al.\,(2004)]
        {pardo04} Pardo, J. R., Alcolea, J., Bujarrabal, V., et al.  2004 A\&A, 424, 145

 \bibitem[Pijpers et al.\,(1994)]
        {pijpers94} Pijpers, F. P., Pardo, J. R., Bujarrabal, V. 1994, A\&A, 286, 501

 \bibitem[Reid\,\&\,Menten\,(1997)] 
        {reid97} Reid, M. J. \& Menten, K. M., 1997, Ap\&\,SS, 251, 41R
        
 \bibitem[Samus et al.\,(2017)]
        {samus17} Samus, N. N., Kazarovets, E. V., Durlevich, O. V., et al. 2017, ARep, 61, 80
   
  \bibitem[Soria-Ruiz et al.\,(2004)]
        {soriaruiz04} Soria-Ruiz, R., Alcolea, J., Colomer. F., et al. 2004, A\&A, 426, 131
    
 \bibitem[Venables and Ripley\,(2002)] 
 {venables02} Venables, W. N \& Ripley, B. D. 2002, Modern Applied Statistics with S, Fourth edition, ed. Springer.
 
\bibitem[Wiesemeyer et al.\,(2009)]
        {wiesemeyer09} Wiesemeyer, H., Thum, C., Baudry, A., Herpin, F. 2009, A\&A, 498, 801

  \bibitem[Willson\,(2000)]
        {willson00} Willson, L. A. 2000, ARA\&A, 38, 573
\end{thebibliography}
\end{document}